\documentclass[usenatbib]{mnras}

\usepackage[T1]{fontenc}
\usepackage{ae,aecompl}

\usepackage{graphicx}	
\usepackage{amsmath}	
\usepackage{amssymb}	

\usepackage[usenames,dvipsnames]{xcolor}
\usepackage{multicol}
\usepackage[normalem]{ulem}

\def \d {\mathrm{d}}
\def \e {\mathrm{e}}
\renewcommand{\vec}[1]{\boldsymbol{\mathbf{#1}}} 
\renewcommand{\vec}[1]{\boldsymbol{\mathbf{#1}}} 
\newcommand{\ms}{\scriptscriptstyle}

\title{On the growth of the thermally modified non-resonant streaming instability}

\author[A. Marret et al.]{
A. Marret$^{1,2,3}$\thanks{E-mail: alexis.marret@obspm.fr},
A. Ciardi$^{2}$,
R. Smets$^{3}$,
J. Fuchs$^{1}$
\\
$^{1}$Ecole Polytechnique, Sorbonne Universit\'e, CNRS, LULI, France \\
$^{2}$Sorbonne Universit\'e, Observatoire de Paris, Universit\'e PSL, CNRS, LERMA, F-75005, Paris, France\\
$^{3}$Sorbonne Universit\'e, Ecole Polytechnique, CNRS, Observatoire de Paris, LPP, F-75005, Paris, France
}

\begin{document}
\label{firstpage}
\pagerange{\pageref{firstpage}--\pageref{lastpage}}
\maketitle

\begin{abstract}
The cosmic rays non-resonant streaming instability is believed to be the source of substantial magnetic field amplification. In this work we investigate the effects of the ambient plasma temperature on the instability and derive analytical expressions of its growth rate in the hot, demagnetized regime of interaction. To study its non-linear evolution we perform hybrid-PIC simulations for a wide range of temperatures. We find that in the cold limit about two-thirds of the cosmic rays drift kinetic energy is converted into magnetic energy. Increasing the temperature of the ambient plasma can substantially reduce the growth rate and the magnitude of the saturated magnetic field.
\end{abstract}



\section{Introduction}
The electromagnetic ion streaming instability occurs when a background plasma is traversed by a population of energetic ions with a drift velocity aligned with the ambient magnetic field. This situation can lead to exponentially growing magnetohydrodynamic-like waves, generated at the expense of the bulk kinetic energy of the streaming particles. Depending on their drift velocity and velocity dispersion, three distinct modes can be excited. In general they grow for streaming velocities larger than the Alfv\'en speed, have a growth time of the order of the ion cyclotron time, and can potentially coexist and compete in their growth. The right-hand resonant mode (RHR, following the nomenclature of \citealt{gary_electromagnetic_1984}) requires a small streaming and thermal velocity and is characterized by magnetic fluctuations
with right-hand polarisation. The left-hand resonant mode (LHR) requires low streaming velocity and large velocity dispersion (\citealt{kulsrud_effect_1969}) and and is left-hand polarized. The non-resonant mode (NR) is right-hand polarized, requires a large drift velocity and its growth is not associated with cyclotron resonances as for the other two modes. The non-resonant mode was first investigated in the context of back-streaming ions from the Earth's bow shock to the foreshock region using a kinetic description (\citealt{sentman_instabilities_1981}, \citealt{winske_diffuse_1984}), and was later derived within a fluid framework and applied to the amplification of magnetic field due to cosmic rays (\citealt{bell_turbulent_2004}). 

Important progress have been made in the last decades to determine whether supernova shocks are able to accelerate cosmic rays up to PeV energy (\citealt{pelletier_turbulence_2006}, \citealt{riquelme_non-linear_2009}, \citealt{ohira_two-dimensional_2009}, \citealt{bai_magnetohydrodynamic-particle--cell_2015}, \citealt{casse_magnetic_2018}, \citealt{crumley_kinetic_2019}). These studies highlight the possible role of the NR mode to amplify the magnetic field fluctuations at supernova shocks at a sufficiently high level for first order Fermi acceleration to take place efficiently. This is in contrast to the two resonant modes which are limited to magnetic field amplification lower than the ambient magnetic field, insufficient to accelerate particles up to PeV energies. However, potentially important damping mechanisms related to the environment where the shock is propagating may also need to be taken into account. Some theoretical studies have started assessing the effects of the ambient medium temperature (\citealt{zweibel_environments_2010}) and collisions with neutrals (\citealt{reville_environmental_2008}), showing that the growth of the NR instability may be reduced in hot and/or collisional environments. 

In this work, we focus on the NR mode and develop the linear theory for the thermally modified instability in the regime where the background protons population is hot and demagnetized. Using hybrid-PIC simulations, we perform a parametric study of the dependency of the magnetic field amplification and saturation over a wide range of background plasma temperature. We show that while the NR mode can generate substantial magnetic field amplification, increasing the background temperature may significantly reduce the growth rate and should be taken into account when modelling cosmic rays acceleration.

This paper is organized as follow: in Sec. \ref{sec:NR_mode}, we first use a simplified fluid approach to capture the fundamental mechanisms driving the growth of the NR instability, and derive the associated spatial and temporal scales. Using a kinetic theoretical framework, we then review the growth rate, real angular frequency and wave number predictions for negligible to small background temperatures, and extend the existing theory to obtain analytical expressions for a hot, demagnetized background plasma. In Sec. \ref{sec:section_simu}, we present the numerical method and the one-dimensional and two-dimensional simulations results. We investigate the growth rate, saturation level, spatial structure, density fluctuations, background plasma heating, and cosmic rays scattering and compare these results to theoretical expectations. Sec. \ref{sec:conclusion} is a summary of the results of this study, and a discussion of possible implications for cosmic rays driven magnetic field amplification scenario.

\section{Modelling the non-resonant mode}
\label{sec:NR_mode}
We consider three populations: a fully ionized background plasma (main protons and electrons, noted with the subscripts '$m$' and '$e$') embedded in a zeroth order magnetic field $\vec B_0=B_0\vec e_x$, and traversed by a population of protons cosmic rays (noted with the subscript '$cr$') with a drift velocity along $\vec B_0$. This plasma is taken to be quasi-neutral: $n_m+n_{cr}=n_e$, where $n_\alpha$ is the density of the species $\alpha$ and initially homogeneous. In order to be consistent with the assumption of an initially homogeneous magnetic field, the total initial current must be null. This is achieved by considering a drift velocity for the electrons population relative to the main protons, in the same direction as the cosmic rays such that:
\begin{align}
\vec{u}_e=\frac{n_{cr}}{n_e}\vec{u}_{cr}
\label{eq:u_e}
\end{align}
A different way of compensating the current would be to distinguish two electrons populations: one with the same density as the main protons, and an additional population with the same charge density as the cosmic rays and drifting alongside them. Within the framework of kinetic theory, \citealt{amato_kinetic_2009} showed that the dispersion relation of the NR mode is only modified by a corrective term of the order $O(n_{cr}^2/n_m^2)$ depending on the choice to compensate the current.

\subsection{Heuristic fluid approach}
\label{sec:mechanism}

To describe the basic mechanism of the NR mode and estimate the characteristic spatial and temporal scales, and the saturated magnetic field associated to the non-resonant mode, we can use a non-relativistic fluid approach. We consider the main protons and electrons as a single fluid with negligible thermal velocity, and investigate the effects of the cosmic rays on this system. Several studies of the instability using this model have been done (\citealt{bell_turbulent_2004}, \citealt{zirakashvili_modeling_2008}, \citealt{bai_magnetohydrodynamic-particle--cell_2015}, \citealt{matthews_amplification_2017}, \citealt{mignone_particle_2018}), and we adopt a similar approach as a starting point for our study.

The momentum conservation equation for the specie $\alpha$, obtained by integration of the Vlasov equation over velocity space, is:
\begin{align}
\rho_\alpha\frac{\d\vec u_\alpha}{\d t}=-\vec\nabla\cdot\vec{P}_\alpha+n_\alpha q_\alpha(\vec{E}+\vec{u}_\alpha\times\vec{B})
\end{align}
In this equation, $\rho_\alpha=n_\alpha m_\alpha$ is the mass density, $\vec u_\alpha$ is the fluid velocity, $\vec{P}_\alpha$ is the pressure tensor, and $\frac{\d}{\d t}=\frac{\partial}{\partial t}+\vec{u}_\alpha\cdot\vec\nabla$ is the material derivative. The collisions have been neglected. By considering non-relativistic velocities, the total current can be expressed with Amp\`ere's law as:
\begin{align}
\vec\nabla\times\vec B = \mu_0 e  (n_m\vec u_m+n_{cr}\vec u_{cr}-n_e\vec u_e)
\end{align}
In this expression, $e$ is the elementary charge and $\mu_0$ the permeability of free space. Performing a summation of the main protons and electrons momentum conservation equations and inserting Amp\`ere's law, one obtains:
\begin{align}
\rho\frac{\d\vec u}{\d t}=-\nabla\cdot\vec{P}+\frac{1}{\mu_0}(\vec\nabla\times\vec B)\times \vec B- en_{cr}(\vec E+\vec u_{cr}\times \vec B)
\label{eq:momentum_conservation}
\end{align}
In Eq. \ref{eq:momentum_conservation}, the background plasma is defined with $\rho= \rho_e+\rho_m$, $\vec u=(\rho_e\vec u_e+\rho_m\vec u_m)/\rho$ and $\vec P=\vec P_{e}+\vec P_{m}$. Note that this plasma is negatively charged to ensure quasi-neutrality. In the following calculation, we will consider $n_{cr}/n_m\ll 1$ such that $n_e\approx n_m$.
The electric field results from the electron Ohm's law, neglecting the electron inertia as well as their pressure. Furthermore, one neglects the Hall effect for spatial scales larger than the protons inertial length, hence obtaining:
\begin{align}
\vec E=-(\vec{u}+\frac{n_{cr}}{n_m}\vec{u}_{cr})\times\vec B
\label{eq:induction}
\end{align}
Neglecting the Hall effect does not hold when considering demagnetized main protons in a collisionless plasma, where the electrons and protons dynamic are not directly correlated. This will be further investigated in Sec. \ref{sec:linear_theory}.

In the following, we will use the reference frame of the initially at rest background fluid. To simplify this heuristic investigation of the instability and to highlight the destabilizing effects of the magnetic force driving term $-en_{cr}\vec u_{cr}\times \vec B$ in Eq. \ref{eq:momentum_conservation}, we neglect the background fluid pressure gradients, and consider the cosmic rays as drifting with a constant and unperturbed velocity $\vec u_{cr}=u_{\parallel cr}\vec e_x$. We also make the assumption of electromagnetic fluctuations with wavelengths smaller than the cosmic rays gyroradius. The resulting Maxwell-Faraday and momentum equations, with a first order linearization of the magnetic and background fluid velocity fluctuations, give:
\begin{align} 
\label{eq:momentum_heur}
\dfrac{\partial\vec u_1}{\partial t} &= \dfrac{(\vec B_0\cdot\vec\nabla)\vec B_1}{\mu_0 \rho} + \dfrac{n_{cr}}{n_m}\Omega_0\left(\vec u_1\times \dfrac{\vec B_0}{B_0} \right) - \dfrac{\vec j_{cr}\times\vec B_1}{\rho} \\
\label{eq:mag_heur}
\dfrac{\partial\vec B_1}{\partial t}&+\vec\nabla\cdot\left(\dfrac{n_{cr}}{n_m}\vec u_{cr}\vec B_1\right)=(\vec B_0\cdot\vec\nabla)\vec u_1
\end{align}
where $\Omega_0=eB_0/m_p$ is the proton cyclotron frequency, $m_p$ is the proton mass and $\vec j_{cr}=en_{cr}\vec u_{cr}$ is the current carried by the cosmic rays. The subscripts '0' and '1' refer to the order for the linearization. 

Many of the underlying features of the instability can be understood by inspecting these equations. The \textit{first term} on the right hand-side of Eq. \ref{eq:momentum_heur} is the magnetic tension force associated to the fluctuating magnetic field and dominates the background fluid dynamic at small enough scale. The \textit{second term} is responsible for a background fluid cyclotron-like motion at a fraction $\omega_u=\Omega_0 n_{cr}/n_m$ of the cyclotron frequency, resulting from the ambient magnetic field and from the excess of negative charge that compensates the cosmic rays charge. The \textit{third term} is the source of the instability and drives growing background fluid velocity fluctuations via the interaction of the cosmic rays current with the fluctuating magnetic field. The linearized magnetic field induction equation (Eq. \ref{eq:mag_heur}) has been rewritten to highlight its conservative character and the presence of a source term, which is unchanged by the presence of cosmic rays and couples the background fluid velocity fluctuations to the magnetic field ones. The plasma being homogeneous, the second term on the left-hand side can be rewritten as a perturbed magnetic field advection term at a velocity $\frac{n_{cr}}{n_m}u_{\parallel cr}$ (equal to the zeroth order electrons current velocity).

To capture quantitatively the mechanism of the non-resonant instability it is only necessary to retain the coupling terms (i.e. neglect the second term in the right-hand side of Eq. \ref{eq:momentum_heur} and the second term in the left-hand side of Eq. \ref{eq:mag_heur}). This is equivalent to supposing fast growing modes, with growth time much smaller than those associated to the perturbed magnetic field advection and to the background fluid cyclotron-like motion. We consider velocity perturbations as $\vec u_1\e^{i(kx-\omega t)}$, and a circularly polarized magnetic perturbation propagating along the $x$ direction such that $\vec B=B_0\vec e_x+\vec B_1\e^{i(kx-\omega t)}$. We have defined the angular frequency $\omega=\omega_r+i\gamma$ where $\omega_r$ is taken to be positive, and the wave number $k$ can be either positive or negative depending on the direction of propagation. Solving Eqs. \ref{eq:momentum_heur} and \ref{eq:mag_heur}, one finds:
\begin{align}
\omega=\left(\dfrac{n_{cr}}{n_m}u_{\parallel cr}\Omega_0k+v_{A0}^2k^2\right)^{1/2}
\label{eq:omega_coupled}
\end{align}
where $v_{A0}=B_0/(\mu_0n_mm_p)^{1/2}$ is the Alfv\'en velocity. 
The second term on the right-hand side of Eq. \ref{eq:omega_coupled} corresponds to the magnetic tension and acts as a stabilizing term by preventing large wave numbers to grow when it is equal or greater to the magnetic force driving term. For $k$ positive, the angular frequency is purely real and the only effect of the cosmic rays is to modify the dispersion of long wavelengths Alfv\'en waves. However for negative $k$ values, we obtain an instability for $|k|<k_{\max}$ with
\begin{align}
k_{\max}=\frac{n_{cr}}{n_m}\frac{u_{\parallel cr}}{v_{A0}^2}\Omega_0
\label{eq:kmax}
\end{align}
The unstable modes propagate backward relative to the CR drift velocity, and with a right-hand polarization (corresponding to a negative helicity, see Appendix \ref{sec:appendix_hel}). Searching for an extremum of Eq. \ref{eq:omega_coupled}, one finds the fastest growing mode as $\gamma=\frac{1}{2}\frac{n_{cr}}{n_m}\frac{u_{\parallel cr}}{v_{A0}}\Omega_0$ and the corresponding wave number $|k|=\frac{1}{2}\frac{n_{cr}}{n_m}\frac{u_{\parallel cr}}{v_{A0}^2}\Omega_0$, which are identical to what can be obtained from kinetic theory calculations in the case of negligible temperatures (see Sec. \ref{sec:linear_theory}).

Thus, in the range of wave numbers $|k|\ll k_{\max}$, the contribution of the first term in the right-hand side of Eq. \ref{eq:momentum_heur} corresponding to the magnetic tension can be neglected, and one obtains that the perturbed background fluid velocity is amplified by the interaction of the cosmic rays current with the perturbed magnetic field. As a result the source term of Eq. \ref{eq:mag_heur} is also amplified, which corresponds to the induced first order electric field $\vec E_1=-\vec u_1\times\vec B_0$ closing the feedback loop by enhancing the magnetic perturbation. This leads to the exponential growth of the electromagnetic wave with a growth rate $\gamma$ varying as $k^{1/2}$:
\begin{align}
\gamma&\approx\left(\dfrac{n_{cr}}{n_m}\Omega_0u_{\parallel cr}|k|\right)^{1/2}
\label{eq:omega_coupled_simplified}
\end{align}
One also find from Eq. \ref{eq:momentum_heur} that the field of velocity fluctuations grows with a phase shift of $-\pi/2$ with respect to the magnetic perturbation. This distinctive geometrical property will be further investigated in Sec. \ref{sec:section_simu} as it is responsible for the development of large anisotropies in the background plasma. We note that the energy exchange between the cosmic rays and the waves is accomplished through the second order parallel electric field $\vec E_{\parallel}=-\vec u_1\times\vec B_1$, which slows down the cosmic rays and accelerates the background plasma. 
 
A lower limit for the unstable wave numbers can be obtained by examining the perturbed magnetic field advection term of Eq. \ref{eq:mag_heur}. The magnetic field perturbation propagates in the direction opposite to the cosmic rays drift velocity. At a given position, this corresponds to a rotation of the magnetic perturbation at a frequency $\omega_B=\frac{n_{cr}}{n_m}u_{\parallel cr}|k|$. In the range $|k|u_{\parallel cr}<\Omega_0$, one has $\omega_B<\omega_{u}$ where $\omega_{u}=\Omega_0 n_{cr}/n_m$ meaning that the driving force $-\vec j_{cr}\times \vec B_1$ is unable to impose the electromagnetic wave frequency to the background fluid motion, which prevents the growth of the magnetic fluctuations. This limit can also be found by considering the time $(|k|u_{\parallel cr})^{-1}$ for cosmic rays to cross one wavelength, which has to be smaller than the cyclotron period $\Omega_0^{-1}$. 
It corresponds to a magnetization condition stopping the exponential growth, as the cosmic rays start following the perturbed field lines at scales comparable to the cosmic rays Larmor radius. Both approaches yield the same condition:
\begin{align}
k_{\min}=\frac{\Omega_0}{u_{\parallel cr}}
\label{eq:kmin}
\end{align}
In the case of $|k|<k_{\min}$, the cosmic rays velocity perturbations cannot be neglected, and the contribution of the $\vec j_{cr}\times\vec B_1$ term in Eq. \ref{eq:momentum_heur} becomes small. 

An estimate of the saturated magnetic field intensity can be found by studying the time evolution of the two limiting wave numbers $k_{\min}$ and $k_{\max}$. During the instability growth, $B$ increases with time and so does the minimum unstable wave number, whereas the maximum wave number decreases. The magnetic field saturation is expected to occur when $k_{\max}=k_{\min}$ (\citealt{bell_turbulent_2004}). This condition can be rewritten in term of energies, and is fulfilled when the magnetic energy equals the kinetic energy of the drifting cosmic rays. The corresponding magnetic field is then estimated by considering the cosmic rays drift velocity to be constant. This yields a saturated magnetic field energy density equal to the initial cosmic rays drift kinetic energy density. For relativistic cosmic rays drift velocities, the $k_{\min}$ limit is expressed as $k_{\min} = \Omega_0/u_{\parallel cr}\gamma_{cr}$ where $\gamma_{cr}$ is the cosmic rays Lorentz factor (\citealt{amato_kinetic_2009}, \citealt{zacharegkas_modeling_2019}); in this case Bell's saturation criterion is written as $B^2/2\mu_0=n_{cr}m_p\gamma_{cr}u_{\parallel cr}^2/2$. In general, depending on the cosmic rays drift kinetic energy, a large magnetic field amplification $B_1>B_0$ can be obtained. This is an important feature of the NR instability, as the RHR and LHR modes are restricted to fluctuations amplification $B_1/B_0\sim 1$ because of the resonance condition on the cosmic rays (\citealt{bell_cosmic_2013}).

Up to this point, we have neglected any potential damping  via thermal effects. In the following section, we will derive the growth rate of the non-resonant mode while taking into account the background protons temperature. We will consider a wide range of parameters, starting from the zero and small temperature regimes (cold and warm plasma) up to the demagnetized regime (hot plasma). We drop the fluid description in order to accurately describe finite Larmor radius effects and focus on obtaining analytical results by expanding the full kinetic dispersion relation for the NR mode.

\subsection{Linear kinetic theory}
\label{sec:linear_theory} 

The kinetic linear dispersion relation for transverse electromagnetic waves ($\vec k\cdot\vec E=0$) propagating in a plasma parallel to an ambient magnetic field is well known (\citealt{scharer_cyclotron_1967}). Considering Maxwellian velocity distribution functions with drift velocities $u_{\parallel\alpha}$, isotropic temperatures $T_{\alpha}$ and $\omega/k \ll c$, where $c$ is the speed of light, the dispersion relation can be written as:
\begin{align}
-k^2c^2-\frac{1}{\sqrt{2}}\sum_\alpha\left[\frac{\omega^2_{p\alpha}}{v_{T\alpha}}\left(u_{\parallel\alpha}-\frac{\omega}{k}\right)Z(\zeta^\pm_\alpha)\right]=0
\label{eq:dispersion}
\end{align}
where the thermal velocity is given by $v_{T\alpha}=(k_BT_{\alpha}/m_\alpha)^{1/2}$, $k_B$ is the Boltzmann constant, $\omega_{p\alpha}=(n_\alpha q_\alpha^2/\varepsilon_0 m_\alpha)^{1/2}$ is the plasma angular frequency, $\Omega_{\alpha}=q_\alpha B_0/m_\alpha$ is the initial cyclotron angular frequency, $\varepsilon_0$ is the permittivity of free space. The summation is performed over all populations $\alpha=e,m,cr$. We introduced the Fried and Conte function (\citealt{fried_plasma_1961}):
\begin{align}
Z(\zeta^\pm_\alpha)=\pi^{-1/2}\int_{-\infty}^{+\infty}\frac{\e^{-u^2}}{u-\zeta_\alpha^\pm}\d u
\end{align}
A key parameter that characterizes the interaction of the population $\alpha$ with the electromagnetic waves of angular frequency $\omega$ and wave number $k$ is the argument of the Fried and Conte functions $\zeta_\alpha^\pm=\dfrac{1}{\sqrt{2}v_{T\alpha}k}(\omega-ku_{\parallel\alpha}+p^\pm\Omega_{\alpha})$, where $p^\pm=+1$ for right-hand polarized waves and $p^\pm=-1$ for left hand polarized waves. Depending on the value of $\zeta_\alpha^\pm$, two regimes of interaction can be distinguished. The first one corresponds to $|\zeta_\alpha^\pm| \gg 1$, where the bulk of the velocity distribution function of population $\alpha$ is far from the cyclotron resonance condition $\omega_r-ku_{\parallel\alpha}+p^\pm\Omega_{\alpha}=0$ (\citealt{gary_second-order_1978}). This so-called cold regime, is non-resonant and may be correctly described using a fluid approach. 
The other regime $|\zeta_\alpha^\pm| < 1$ can be defined as hot and demagnetized, as the thermal Larmor radius is larger than the wavelength of the mode.

\paragraph*{Cold regime $v_{Tm}\to 0\ \ kr_{Lm}<1$} \hfill \break
The expression for the fastest growth rate $\gamma_\mathrm{\ms cold}$ and it's associated real angular frequency $\omega_\mathrm{\ms r,cold}$ and wave number $k_\mathrm{\ms cold}$ for the non-resonant mode were first derived in the cold plasma limit for all populations by \citealt{winske_diffuse_1984}, using a non-relativistic kinetic framework and considering protons populations with a small density ratio $n_{cr}/n_m$. In this limit, Eq. \ref{eq:dispersion} can be simplified using asymptotic expansions of the Fried and Conte function, and one finds:
\begin{align} 
\gamma_\mathrm{\ms cold} = &\ \frac{1}{2}\frac{n_{cr}}{n_{m}}\frac{u_{\parallel cr}}{v_{A0}}\Omega_{0} \label{eq:gamma} \\
\omega_\mathrm{\ms r,cold} =& \frac{1}{2}\left(\frac{n_{cr}}{n_m}\frac{u_{\parallel cr}}{v_{\ms A0}}\right)^2\Omega_0 \label{eq:omeg} \\
k_\mathrm{\ms cold} =&\frac{1}{2}\frac{n_{cr}}{n_m}\frac{u_{\parallel cr}}{v_{A0}^2}\Omega_0 
\end{align}
We have $\gamma_\mathrm{\ms cold}=k_\mathrm{\ms cold}v_{A0}$. Note that the fastest growing mode $k_\mathrm{\ms cold}$ is half of the maximum one $k_{\max}$ (Eq. \ref{eq:kmax}). The growth rate and wave number are identical to those found by \citealt{bell_turbulent_2004} using a fluid description for the background plasma and a power law with relativistic velocities for the streaming population. Both derivations were performed assuming low frequency modes and no resonant interactions. This regime of interaction is particularly relevant in the context of supernova shocks in the interstellar medium and of backstreaming populations from the earth bow shock region (\citealt{onsager_interaction_1991}, \citealt{akimoto_nonlinear_1993}), where thermal effects are expected to be small.

\paragraph*{Warm regime $v_{Tm}\neq 0\ \ kr_{Lm}<1$} \hfill \break
The warm regime corresponds to the limit of finite main protons thermal velocity $v_{Tm}$, but $kr_{Lm}<1$ such that $|\zeta_m^\pm|\gg 1$. In the same way as in the cold regime, Eq. \ref{eq:dispersion} can be simplified using asymptotic expansions but retaining additional terms to account for thermal corrections. The fastest growing mode $\gamma_\mathrm{\ms warm}$ (\citealt{reville_environmental_2008}) and associated wave number $k_\mathrm{\ms warm}$ (\citealt{zweibel_environments_2010}) in the warm regime are found to be:
\begin{align}
\gamma_\mathrm{\ms warm}=&\left(\frac{n_{cr}}{n_m}\frac{u_{\parallel cr}}{v_{Tm}}\right)^{2/3}\Omega_0 \label{eq:gamma_reville} \\
k_\mathrm{\ms warm}=&\left(\frac{n_{cr}}{n_m}\frac{u_{\parallel cr}}{v_{Tm}}\right)^{1/3}\frac{\Omega_0}{v_{Tm}} \label{eq:k_zweibel}
\end{align}
The growth rate in this regime depends linearly on the initial magnetic field, and as $T_m^{-1/3}$. Finite Larmor radius effects tend to reduce the NR mode growth and shift the unstable wavelengths toward larger scales. A threshold for this regime can be calculated as $v_{\ms{A0}}/v_{Tm}<(n_{cr}u_{\parallel cr}/n_mv_{Tm})^{1/3}$. The warm regime of interaction is of interest in low density, high temperature medium such as superbubbles, where the non-resonant mode may be significantly damped. 

\paragraph*{Hot regime $v_{Tm}\neq 0\ \ kr_{Lm}>1$} \hfill \break
Following the analysis of \citealt{reville_environmental_2008} and \citealt{zweibel_environments_2010}, we now derive the expressions of the growth rate, real angular frequency, wave vector and phase velocity for the hot, demagnetized regime of interaction $kr_{Lm} > 1$. We present here the results of the calculation, more details can be found in appendix \ref{sec:appendix_linear}. 

\begin{figure}
	\includegraphics[width=\columnwidth]{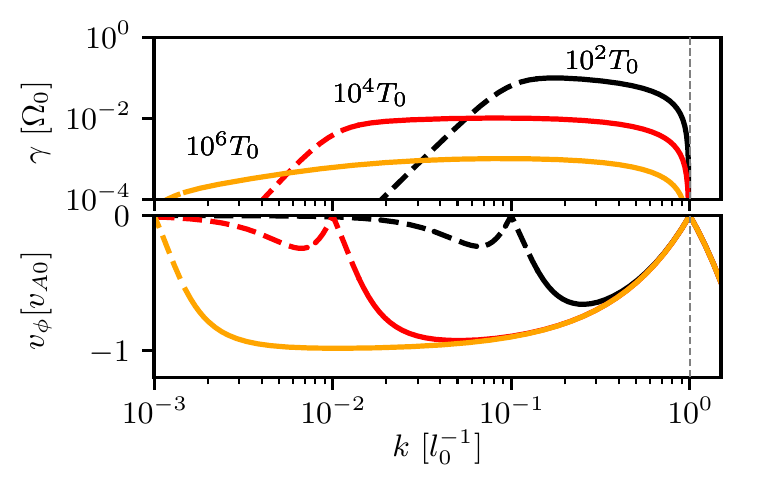}
    \caption{Growth rate $\gamma_\mathrm{\ms{hot}}$ (upper panel) and phase velocity $v_{\phi}=\omega_\mathrm{r,hot}/k$ (lower panel) as a function of the wave number $k$, obtained from Eqs. \ref{eq:gamma_hot} and \ref{eq:omega_hot}. Parameters used are, in normalized units: $n_{cr}=0.01\ n_m$, $u_{\parallel cr}=100\ v_{A0}$. The black, red and orange curves corresponds to $T_m=10^2,\ 10^4,\ 10^6\ T_0$ respectively. The dotted lines correspond to wave numbers where the demagnetized main protons assumption is not fulfilled. The grey vertical dotted line corresponds to $k=k_{\max}$ from Eq. \ref{eq:kmax}.} 
    \label{fig:figure_1}
\end{figure}

We restrict ourselves to low frequency waves, such that $\omega<\Omega_{\alpha}$, and consider a single electron population (whereas two populations were considered in \citealt{zweibel_environments_2010}) and a Maxwellian distribution for the cosmic rays (whereas a mono-energetic distribution was considered in \citealt{reville_environmental_2008}). These differences have no impact on the final results as long as the density ratio $n_{cr}/n_m$ is small before unity and the electrons are magnetized. Neglecting electron inertia (which is equivalent to the low frequency assumption), we obtain the expressions of the thermally modified growth rate $\gamma_\mathrm{\ms{hot}}(k)$ and real angular frequency $\omega_\mathrm{\ms{r,hot}}(k)$ for the hot regime of the non-resonant streaming instability:
\begin{align}
\thinmuskip=0mu
\medmuskip=0mu
\thickmuskip=0mu
\label{eq:gamma_hot}
\begin{split}
\gamma_\mathrm{\ms{hot}}(k) =  
\dfrac{(2\pi)^{1/2}}{r_{Lm}\xi}
\dfrac{\dfrac{k}{\Omega_0}\biggl(v_{A0}^2-\dfrac{n_{cr}}{n_m}u_{\parallel cr}^2\biggr)-p^\pm\biggl(\dfrac{k^2}{\Omega_0^2}v_{A0}^2+\dfrac{n_{cr}^2}{n_m^2}\biggr)u_{\parallel cr}}{\dfrac{\pi}{k^2r_{Lm}^2}+2\biggl(\dfrac{1}{k^2r_{Lm}^2}-\dfrac{n_{cr}}{n_m}\dfrac{1}{\xi}-1\biggr)^2}
\end{split}
\end{align}
\begin{align}
\label{eq:omega_hot}
\omega_\mathrm{\ms{r,hot}}(k) = \dfrac{k^3r_{Lm}^2\left(k^2r_{Lm}^2-1\right)\left(\dfrac{n_{cr}}{n_m}u_{\parallel cr}+p^\pm \dfrac{k}{\Omega_0}v_{A0}^2\right)}{k^4r_{Lm}^4+k^2r_{Lm}^2\left(\dfrac{\pi}{2}-2\right)+1}
\end{align}
where we have defined the parameter $\xi=p^\pm ku_{\parallel cr}/\Omega_0-1$. The growth rate $\gamma_\mathrm{\ms{hot}}(k)$ and phase velocity $v_\phi=\omega_\mathrm{\ms{r,hot}}(k)/k$ are plotted in Fig. \ref{fig:figure_1} for the main protons temperatures $T_m =10^2, 10^4, 10^6\ T_0$ (see Sec. \ref{sec:num_petup} for a discussion on the normalization).
The growth rate is found to be strongly reduced with increasing temperature, and the fastest growing mode shifts towards smaller wave numbers compared to the cold regime. In the warm regime, finite Larmor radius effects of the main protons play a role in determining the largest unstable wave number. We find that in the hot regime however, the competition between the magnetic tension and the cosmic rays current  driving term is the only determining factor of the largest unstable wave number, and one obtains good agreement with the fluid estimate $k_{\max}=\frac{n_{cr}}{n_m}\frac{u_{\parallel cr}}{v_{A0}^2}\Omega_0$. This can be understood by considering the fluid model presented in Sec. \ref{sec:mechanism} while retaining the Hall effect in Ohm's law (Eq. \ref{eq:induction}) to account for the decoupling between electrons and  background protons in the demagnetized and collisionless regime. One then finds that for all the unstable wavelengths, the resulting background fluid momentum conservation equation is not modified, resulting in identical maximum unstable wavenumber $k_{\max}$ in both the cold and hot regimes.

Useful analytical expressions can be obtained by considering the limits $kr_{Lm}\gg 1$, $ku_{\parallel cr}/\Omega_0\gg 1$ which corresponds to the hypothesis of demagnetized main protons, and to the instability requirement $|k|>k_{min}$ discussed in Sec. \ref{sec:mechanism}. One finds the approximate expressions for the fastest growing mode:
\begin{align}
\gamma_\mathrm{\ms{hot}} &= \left(\frac{\pi}{2}\right)^{1/2}\frac{n_{cr}}{n_m}\frac{u_{\parallel cr}}{v_{Tm}}\Omega_0 \label{eq:gamma_hot_analytic}\\
\omega_\mathrm{\ms{r,hot}} &= \frac{n_{cr}^2}{n_m^2}\frac{u_{\parallel cr}}{v_{A0}}\Omega_0  
\label{eq:omega_hot_analytic}\\
k_\mathrm{\ms hot}&=\frac{n_{cr}}{n_m}\frac{\Omega_0}{v_{A0}} 
\label{eq:k_hot_analytic}\\
v_\mathrm{\phi,\ms hot}&=-\frac{n_{cr}}{n_m}u_{\parallel cr}
\label{eq:vphi_hot_analytic}
\end{align}
We give here the absolute value of $\omega_\mathrm{\ms{r,hot}}$ and $k_\mathrm{\ms hot}$. For $\zeta_m^\pm\gtrsim 1/2$, the first order asymptotic expansion of the main protons Fried and Conte function cannot accurately describe the complete function. As a consequence, Eqs. \ref{eq:gamma_hot_analytic} to \ref{eq:vphi_hot_analytic} hold for $k_\mathrm{\ms cold}r_{Lm}\gtrsim 2$ which corresponds to the demagnetization of half of the fastest growing mode in the cold limit. 

The growth rate for hot, demagnetized main protons is found to decrease as $T^{-1/2}$ with temperature, more rapidly than the $T^{-1/3}$ dependency in the warm protons regime, and we obtain $\gamma_\mathrm{\ms hot}/\gamma_\mathrm{\ms cold} = (2\pi)^{1/2}v_{A0}/v_{Tm}$. This result may be of importance in high temperature plasmas with small ambient magnetic field, where the instability growth may be strongly reduced. We find that the real angular frequency and the fastest growing wave number are independent of the main protons temperature, and the fastest growing wave number is also independent of the cosmic rays velocity. The phase velocity $v_\mathrm{\phi,hot}=\omega_\mathrm{r,hot}/k_\mathrm{hot}$ is equal and opposed to the electron drift velocity compensating the cosmic rays current, which is the same result as in the cold regime. We will return to these results in Sec. \ref{sec:conclusion} where we discuss possible applications for astrophysical settings.

Having studied the instability linear theory for a large range of temperature, we will now use hybrid-PIC simulations to verify the theory developed in the last two sections, and explore the non-linear behaviour of the unstable waves. We will first present our numerical model, then our 1D and 2D simulations results.

\section{Simulations results}
\label{sec:section_simu}
We use the Hybrid-PIC code HECKLE (\citealt{smets_r_heckle_2020}), which solves the Vlasov-Maxwell system using a predictor-corrector scheme for the electromagnetic field and a non-relativistic Boris pusher (\citealt{boris_acceleration_1970}) for the particles. The main and cosmic rays protons are described as macro-particles, and the electrons as a mass-less fluid. This hybrid approach is well suited to study the kinetic, non-linear evolution of systems at the protons temporal scale while avoiding prohibitive computational time.
\subsection{Numerical model and setup}
\label{sec:num_petup}

Masses and charges are normalized to the proton mass $m_p$ and elementary charge $e$ respectively. The densities and magnetic field are normalized to a reference value $n_0=n_m(t\!=\!0)$ and $B_0 = B(t\!=\!0)$. Frequencies, lengths and velocities are normalized to the initial proton cyclotron angular frequency $\Omega_0=eB_0/m_p$, initial proton inertial length $l_0=c/\omega_{pm}$ where $c$ is the speed of light, $\omega_{pm}=(n_0e^2/\varepsilon_0 m_p)^{1/2}$ is the protons plasma frequency and $v_{A0}=B_0/(\mu_0n_0m_p)^{1/2}=l_0\Omega_0$ is the initial Alfv\'en velocity. Temperatures are expressed in units of energy as $T_0=m_pv_{A0}^2$. The motion of a macro-particle $k$ is obtained as:
\begin{align}
\frac{\d \vec v_k}{\d t} = \frac{q_k}{m_k}(\vec E+\vec v_k\times\vec B)
\end{align}
The electric field $\vec{E}$ is normalized to $E_0=v_{A0}B_0$. Maxwell's equations are solved in the non-relativistic regime:
\begin{align}
\frac{\partial \vec B}{\partial t} = -\vec\nabla\times\vec E
\end{align}
\begin{align}
\vec J = \vec\nabla\times\vec B/\mu_0
\end{align}
Note that the current $\vec J$ in the Hall term is only the transverse one. Quasi-neutrality is assumed at each time step. The electric field is computed using the generalized Ohm's law:
\begin{align}
\vec E = -\vec u_i\times\vec B+\frac{1}{en_e}(\vec J\times\vec B-\vec\nabla\cdot\vec P_e)+\sigma\vec J-\sigma'\Delta\vec J
\end{align}
where $\sigma$ is the resistivity, and $\sigma'$ the hyperviscosity. These coefficients are taken to be $10^{-3}B_0/e n_0$ and $10^{-3}B_0l_0^2/en_0 $ respectively in order to reduce small scale fluctuations without introducing important dissipative effects. Electron inertia terms have been neglected ($m_e=0$), consistent with the long time scale assumption. $\vec u_i$ and $en_e$ are the fluid velocity and charge density calculated over the ions populations $l$ as:
\begin{align}
en_e(\vec x)=&\textstyle\sum_{\ell,k}q_\ell W_\ell S(\vec x-\vec x_{\ell,k}) \\
\vec u_i(\vec x)=&\textstyle\sum_{\ell,k}\vec v_{\ell,k}W_\ell S(\vec r-\vec x_{\ell,k})/\textstyle\sum_{\ell,k}W_\ell S(\vec r-\vec x_{\ell,k})
\end{align}
In these expressions, $\vec x$ is the grid point position, $\vec x_{l,k}$ the position of a macro-particle $k$ from population $l$, and $S(\vec x-\vec x_{\ell,k})$ the first order B-spline. Ions populations have different numerical weights $W_\ell$ for fluid quantities calculation, allowing the simulation of different densities while keeping the same number of macro-particles for each species. The electron pressure is calculated by supposing an isothermal, isotropic behavior:
\begin{align}
P_e=n_ek_BT_e
\end{align}
where $k_B$ is the Boltzmann constant, and $T_e$ the uniform electron temperature fixed at the beginning of the simulation.

The simulations are performed in 1D and 2D space. Vector quantities are defined in 3D. We consider two populations of protons with an initial Maxwellian velocity distribution function. The ambiant magnetic field is initially homogeneous and oriented in the $x$ aligned with the simulation domain. The cosmic rays population of density $n_{cr}=0.01\ n_0$ is given a positive drift velocity parallel to the ambient magnetic field $u_{\parallel cr}= 100\ v_{A0}$, in the reference frame of the main protons. The electron density and initial velocity are calculated to ensure quasi-neutrality and satisfy the initial current condition (Eq. \ref{eq:u_e}). In this configuration, unstable waves are expected to propagate with a negative phase velocity, right-hand polarization and negative helicity. 

We use a simulation domain of length $L_x=1000$ $l_0$ discretized with 1000 cells for one-dimensional simulations. These dimensions are  sufficient to simulate the expected range of unstable wave numbers $k_{\max}=l_0^{-1}$ and $k_{\min}=0.01\ l_0^{-1}$ obtained from Eqs. \ref{eq:kmax} and \ref{eq:kmin}, and to correctly model cascade effects. The plasma and field quantities are initially homogeneous. The time step is fixed at $10^{-4}$ $\Omega_0^{-1}$ to properly satisfy the CFL condition on the whistler waves and the most energetic macro-particles. We use 1000 macro-particles per cell initially (500 for each proton populations) to properly describe high temperature Maxwellian distributions, as well as the large density fluctuation that occur during the instability growth. For two-dimensional simulations, we use a domain length $L_y=200\ l_0$ in the y-direction discretized with 200 cells. We also performed simulations with $L_y=400$ $l_0$ discretized with 400 cells, without any noticeable changes in the results. Periodic boundary conditions are used in all directions. Collisions are not considered. The simulations setup is an initial value problem as the cosmic rays population is not injected over time during the simulation. We choose $T_e$ and $T_{cr}$ to be equal to the reference temperature $T_0$ for the electrons and cosmic rays, and focus on studying initial main protons temperatures in the range $T_m=0.1$ to $200\ T_0$. A summary of the simulation parameters can be found in Table \ref{table:parameters_simulations}. 

\begin{table*}
\begin{center}
\begin{tabular}{| c | c | c | c | c | c | c | c | c | c | c | c |} \hline
dim & $n_{cr}/n_m$ & $u_{\parallel cr}$ & $\beta_m$ & $T_m$ & $T_{cr}$ & $T_e$ & $L_x$/$L_y$ & $\Delta x$/$\Delta y$\\\hline
1D & 0.01 & 100 & 0.2 to 400 & 0.1 to 200 & 1 & 1 & 1000/$\ \ \ \ $ & 1/$\ $ \\
2D & 0.01 & 100 & 20, 50, 100 & 10, 25, 50 & 1 & 1 & 1000/200 & 1/1 \\ \hline
\end{tabular}
\caption{Normalized parameters used in the simulations. We defined $\beta_m=2(v_{Tm}/v_{A0})^2$. $\Delta x$ and $\Delta y$ are the mesh size in the $x$ and $y$ directions.} 
\label{table:parameters_simulations}
\end{center}
\end{table*}
We will compare simulations with an initial main protons temperature $T_m=T_0$ and $T_m=25\ T_0$ to highlight the change of behaviour of the instability from the cold regime to the warm/hot regimes of interaction. 2D Simulations with $T_m=25\ T_0$ are compared to 1D simulations to retrieve additional informations on the spatial structure of the instability in the hot background plasma limit.

\subsection{Magnetic field amplification}
\label{sec:magnetic_growth}
\begin{figure}
	\includegraphics[width=\columnwidth]{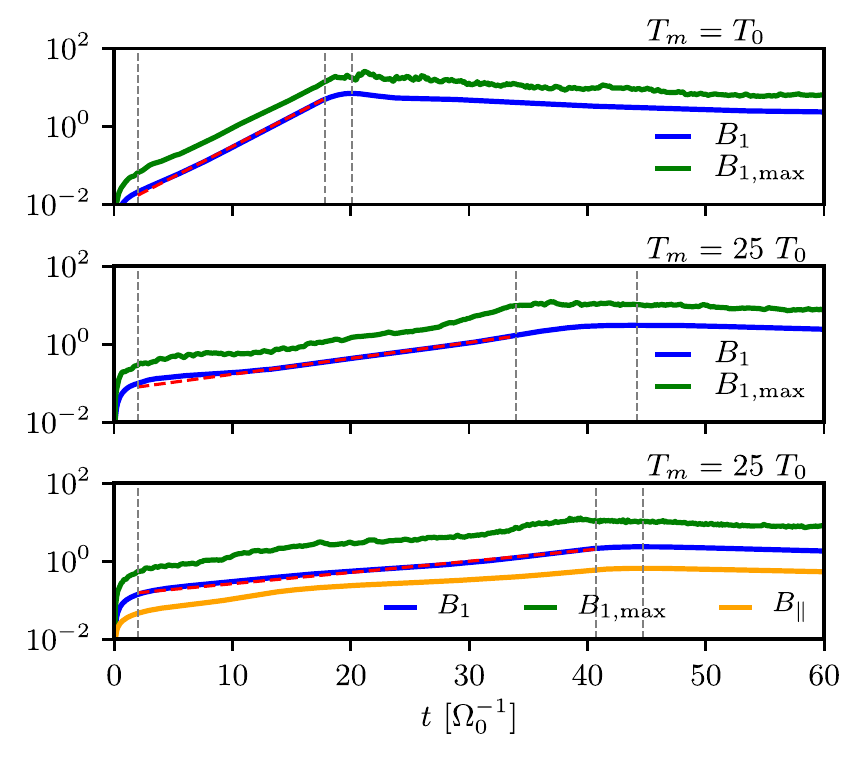}
    \caption{Perturbed magnetic field intensity $B_1=||\vec{B}-\vec{B}_0||/B_0$ evolution over time  integrated over space (blue solid line) and maximum value in simulation domain $B_{1,\max}/B_0$ (green solid line), for 1D simulations with a main protons temperature $T_m=T_0$ (upper panel) and $T_m=25\ T_0$ (middle panel). 2D simulation with $T_m=25\ T_0$ is presented in the lower panel. The red dashed line corresponds to an exponential fit in the linear phase. The orange line in the lower panel correspond to the perturbed magnetic field parallel component $B_\parallel=\vec B_1\cdot\vec e_x$. The vertical dashed lines corresponds, from left to right, to the beginning of the linear regime, transition to the non-linear regime and to magnetic saturation, which is reached typically after 6 e-foldings of growth.}
    \label{fig:figure_2}
\end{figure}

One of the main feature of the non-resonant streaming instability is the generation of large magnetic fluctuations. Contrary to the right-hand and left-hand resonant mode, amplification beyond the initial magnetic field intensity is possible because of the large drift velocity required to drive the instability which keeps the cosmic rays demagnetized (\citealt{bell_cosmic_2013}). The time evolution of the normalized perturbed magnetic field intensity $B_1=||\vec{B}-\vec{B}_0||/B_0$ is presented in Fig. \ref{fig:figure_2} for two different main protons temperatures $T_m=T_0$ and $T_m=25\ T_0$. We highlight four distinct phases. The first one (from $t$ = 0 to 2 $\Omega_0^{-1}$) is purely numerical and corresponds to micro-adjustments of the plasma from the random initialization to its eigenmode values. The second phase is characterized by the exponential growth of the perturbed magnetic field intensity, with a reduced growth rate for larger temperature. A non-linear phase occurs after a few e-foldings of growth, further increasing the magnetic field until the saturation is reached. Finally, the fourth phase corresponds to a slow relaxation of the system with enhanced wave activity. Despite the significant growth of a $B_x$ component in 2D (not permitted in 1D geometry), the growth rate and saturation level are very comparable in 1D and 2D. The simulations yield strong, non-linear amplification of the magnetic field reaching peak values 10 times the ambient magnetic field. In the following, we will focus on the linear and non-linear phases which are of most interest to study the instability behaviour and temperature dependency. 

\begin{figure}
	\includegraphics[width=\columnwidth]{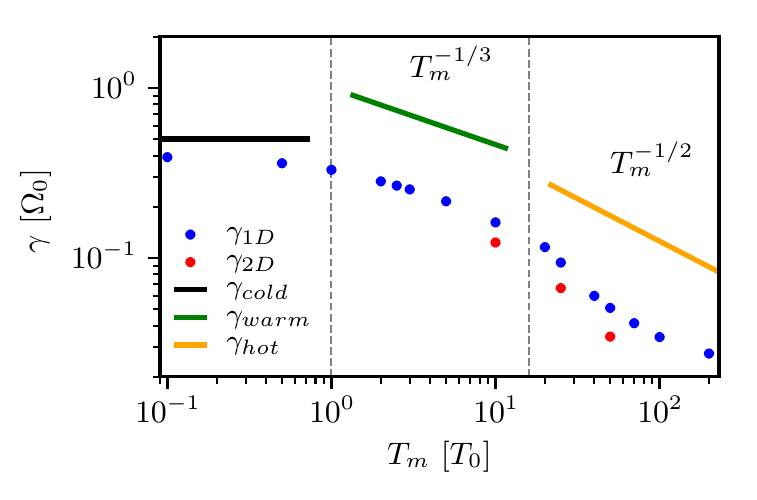}
    \caption{1D and 2D simulations growth rate $\gamma_{\ms{1D,2D}}$ (blue and red dots), \citealt{winske_diffuse_1984} prediction $\gamma_{\ms{cold}}$ (Eq. \ref{eq:gamma}, solid black line), \citealt{reville_environmental_2008} prediction $\gamma_{\ms{warm}}$ (Eq. \ref{eq:gamma_reville}, solid green line) and growth rate prediction of this work $\gamma_{\ms{hot}}$ (Eq. \ref{eq:gamma_hot_analytic}, solid orange line) as a function of the main protons temperature $T_m$. The vertical dashed lines indicates the transition to the warm regime $v_{\ms{A0}}/v_{Tm}<(n_{cr}u_{\parallel cr}/n_mv_{Tm})^{1/3}$ (\citealt{zweibel_environments_2010}, left line at $T_m=T_0$), and to the hot regime $k_{\ms cold}r_{Lm}>2$ (Eq. \ref{eq:kmax}, right line at $T_m= 16\ T_0$). Parameters used are, in normalized units: $n_{cr}=0.01\ n_m$, $u_{\parallel cr}=100\ v_{A0}$.} 
    \label{fig:figure_3}
\end{figure}

One important parameter characterizing the linear phase is the growth rate of the instability. Fig. \ref{fig:figure_3} shows the predictions of the fastest growing mode in the three regimes of cold (Eq. \ref{eq:gamma}), warm (Eq. \ref{eq:gamma_reville}) and hot (Eq. \ref{eq:gamma_hot_analytic}) main protons, alongside growth rates extracted from 1D and 2D simulations $\gamma_{\ms{1D,2D}}$, as a function of the main protons temperature. The growth rate in the hot regime is found to decrease with the temperature as $T_m^{-1/2}$ as expected from the linear theory calculation of this work. In the low temperature limit, the prediction from \citealt{winske_diffuse_1984} is very accurate, and become rapidly invalid for temperatures $T_m>T_0$. The intermediate warm regime from $T_0$ to $16\ T_0$ is well reproduced by the prediction from \citealt{reville_environmental_2008} with a decrease of the growth rate with temperature as $T_m^{-1/3}$. The overestimates in the warm and hot regimes by a factor $\sim 2$ may be linked to the fact that the theoretical values correspond to the fastest growing mode. The magnetic field intensity in the simulations is integrated over the whole $k$ spectrum, which gives an overall smaller growth rate than if only the fastest growing mode was observed.

\begin{figure}
	\includegraphics[width=\columnwidth]{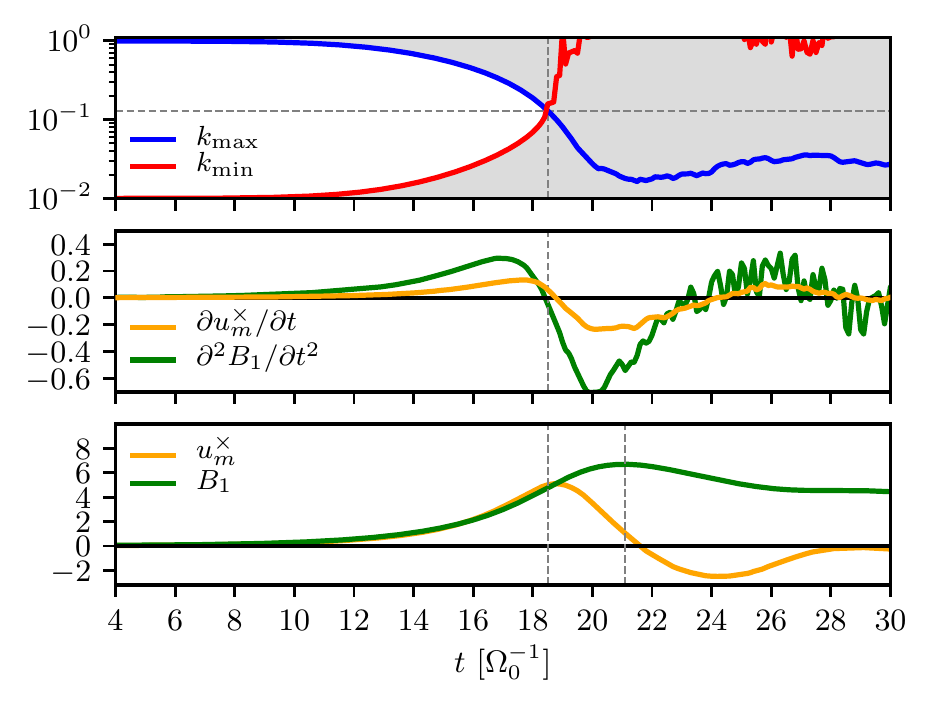}
    \caption{Upper panel: maximum (blue solid line) and minimum (red solid line) unstable $k$ (in unit of $l_0^{-1}$), as a function of time between $t=4\ \Omega_0^{-1}$ and $t=30\ \Omega_0^{-1}$. The condition $k_{\max}=k_{\min}$ is indicated with the vertical dashed line at $t_{NLT}= 18.5\ \Omega_0^{-1}$ and reported in other panels. Greyed regions correspond to stable wave numbers. Middle panel: first order derivative over time of the main protons normal velocity $u_m^\times$ (in unit of $v_{A0}$, orange solid line) and perturbed magnetic field intensity second order derivative over time (in unit of $B_0$ and multiplied by a factor 100, green solid line). Lower panel: perturbed magnetic field intensity $B_1$ (green solid line) and main protons normal fluid velocity (orange solid line). Magnetic saturation is indicated with the vertical dashed line at $t_{sat}=21\ \Omega_0^{-1}$. Values are taken from 1D simulation with a main protons temperature $T_m=T_0$.}
    \label{fig:figure_4}
\end{figure}

To study the instability behaviour during the linear phase, we investigate the time evolution of the maximum and minimum unstable wave numbers $k_{\max}=\frac{n_{cr}}{n_m}\frac{u_{\parallel cr}}{v_{A0}^2}\Omega_0$ and $k_{\min}=\Omega_0/u_{\parallel cr}$, which are expected to play a central role in determining the maximum reachable magnetic field. The results are presented in Fig. \ref{fig:figure_4} upper panel. As the magnetic field increases, $k_{\max}$ decreases whereas $k_{\min}$ increases. The moment these wave numbers become equal corresponds to the magnetic field saturation proposed in \citealt{bell_turbulent_2004}. The instability condition $k_{\min}<|k|<k_{\max}$ cannot be satisfied at any scale, and one expect to obtain a decrease of the main protons velocity in the $(\vec e_y,\vec e_z)$ plane.

To quantify the effects of the instability on the velocities of the proton populations, we will use a local magnetic field aligned basis $\vec e_\parallel=\frac{\vec B_0}{B_0}=\vec e_x$ (parallel component), $\vec e_\times=\frac{\vec B_1}{B_1}\times\vec e_\parallel$ (normal component) and $\vec e_\perp=\vec e_\parallel\times\vec e_\times$ (perpendicular component, aligned with the perturbed magnetic field for an electromagnetic wave propagating along $\vec{B}_0$). As this vector basis is built to follow the local magnetic perturbation, the spatial average of any quantities on this frame of reference does not create any loss of information on the periodic space dependency of the wave. Fig. \ref{fig:figure_4} middle panel presents the derivative over time of the main protons normal fluid velocity component $u^\times_m$, which corresponds to the direction of application of the magnetic force in the local magnetic field basis. The acceleration is increasing exponentially during the linear phase, starts to decrease after $t=17\ \Omega_0^{-1}$, and then becomes negative, corresponding to a slowing down of the main protons rotation. The fluctuating magnetic field second order derivative over time, expected to be closely related to the velocity field (Eq. \ref{eq:mag_heur}), is also shown and exhibits the same behaviour, confirming the correlation between the main protons fluid motion and the growth of the magnetic perturbation. One obtain an excellent match between the $k_{\max}=k_{\min}$ condition discussed previously and the deceleration of the main protons velocity. This suggests that this condition is correlated to the transition toward a non-linear phase, and not to magnetic saturation as the magnetic field keeps growing, although at a slower rate. We recover the same correlation for all our simulations, indicating that the $k_{\max}=k_{\min}$ condition may be a robust criteria to identify quantitatively the end of the exponential growth. We note that linear theory describes very well the instability growth even for large magnetic perturbation as the non-linear transition occurs when the perturbed magnetic field intensity is already greater than the initial ambient magnetic field.

The non-linear phase which follows the linear phase of the instability is characterized by a decrease of the main protons fluid rotation velocity and a reduced magnetic field growth. Fig. \ref{fig:figure_4} lower panel presents the main protons normal velocity and perturbed magnetic field intensity evolution over time. The transition toward non-linear growth, correlated to the maximum in normal velocity $u_{m}^{\times}$ is shown with the vertical dashed black line at $t_{NLT}= 18.5\ \Omega_0^{-1}$, and the magnetic field saturation by the second vertical dashed black line at $t_{sat}=21\ \Omega_0^{-1}$ corresponding to the maximum in magnetic field intensity. The magnetic field keeps growing during the non-linear phase until the normal velocity component becomes negative, corresponding in the magnetic field aligned basis to a loss of the $-\pi/2$ phase shift with respect to the magnetic perturbation necessary to the growth of the NR mode, as expected from the fluid model of the instability presented in Sec. \ref{sec:mechanism}. As a consequence, the parallel induced electric field changes sign and no longer slows down the cosmic rays drift velocity (Eq. \ref{eq:induction}), leading to the magnetic field saturation. This saturation mechanism is well observed in all our simulations. The normal velocity component decrease during the non-linear phase is due both to the conversion of the remaining rotational kinetic energy accumulated during the linear phase into magnetic energy via the induced electric field, and to the loss of the coupling between the magnetic perturbation and the main protons fluid rotation as the magnetic force driving term no longer operates, which leads to a decrease of the normal velocity component (and an increase of the perpendicular one) in the local magnetic field aligned basis. 

The saturated magnetic field intensity is a key parameter of the instability in the context of supernova shocks, as it dictates whether cosmic rays can be confined and accelerated via first order Fermi acceleration. As discussed in Sec. \ref{sec:mechanism}, the fluid model predicts that at saturation the magnetic energy density equals the cosmic rays drift kinetic energy density. An estimate for the saturated magnetic field can then be found by assuming the cosmic rays to be drifting with a constant velocity (\citealt{bell_turbulent_2004}). A different estimate can be found by considering energy exchange rates within quasi-linear theory calculations (\citealt{winske_diffuse_1984}, \citealt{winske_electromagnetic_1986}), which yield that the rate of energy gained by the magnetic field is half of the rate of loss of the cosmic rays drift kinetic energy. Extrapolating this result to saturation and supposing that the cosmic rays drift velocity is null at saturation, one obtains for the magnetic energy density:
\begin{equation}
\dfrac{B^2}{2\mu_0}\sim\dfrac{1}{4} n_{cr}m_pu_{\parallel cr}^2
\label{eq:saturation}
\end{equation}
which is half of the fluid prediction obtained from the condition $k_{\min}=k_{\max}$.

However, kinetic theory calculations show that for the instability to exist, the cosmic rays drift velocity must be larger than the Alfv\'en speed in the amplified field (\citealt{gary_electromagnetic_1984}). In some regimes, this condition is violated and the growth of the instability is halted before the $k_{\min}=k_{\max}$ limit is reached (\citealt{riquelme_non-linear_2009}). All the difficulty lies in assessing the highly non-linear evolution of the cosmic rays drift velocity, which would then determine whether the conditions $k_{\min}=k_{\max}$ or $u_{\parallel cr}\sim v_A$ will give the most accurate saturation mechanism, and whether the assumption of constant or completely depleted drift kinetic energy is relevant to estimate the saturated magnetic field. As such, only numerical simulations can provide a precise answer.

\begin{figure}
	\includegraphics[width=\columnwidth]{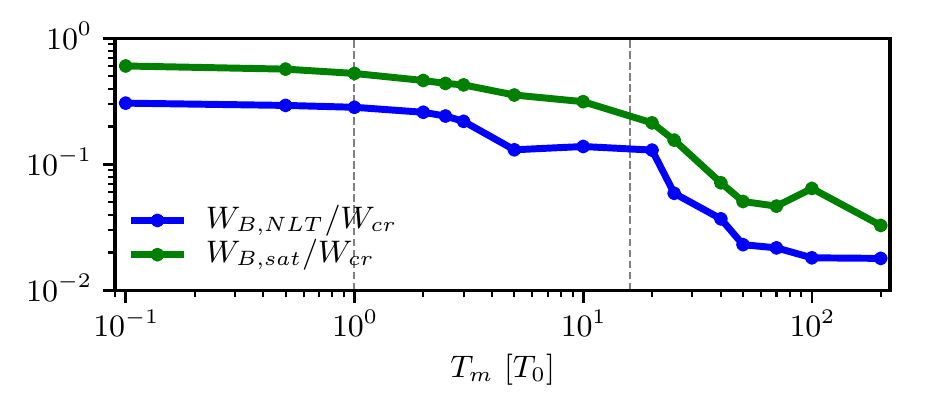}
    \caption{Magnetic field energy density $W_B=B^2/2\mu_0$ normalized to the initial cosmic rays drift kinetic energy density $W_{cr}=n_{cr}m_pu_{\parallel cr}^2/2\ (t_0)=50\ l_0^{-3}m_pv_{A0}^2$ (see Table \ref{table:parameters_simulations}), as a function of the main protons temperature for 1D simulations. Blue curve corresponds to the ratio at non-linear transition (noted $NLT$) and green curve to the ratio at magnetic saturation (noted $sat$). The non-linear transition time is found numerically by equating $k_{\max}=k_{\min}$ averaged in the simulation box. The two dashed vertical lines corresponds to the limits of the warm and hot regimes of interaction as in Fig. \ref{fig:figure_3}.}
    \label{fig:figure_5}
\end{figure}

Fig. \ref{fig:figure_5} presents the ratio between the magnetic field energy density $W_B=B^2/2\mu_0$ and the initial cosmic rays kinetic energy density $W_{cr}=n_{cr}m_pu^2_{\parallel cr}/2$, at non-linear transition (blue solid line) and at saturation (green solid line), as a function of the main protons temperature. In the cold regime, the simulations yield a conversion efficiency of 30 per cent at the transition from linear to non-linear growth, and about 60 per cent at saturation which is close to the quasi-linear theory prediction. The intermediate, warm regime of interaction shows a quick decrease of the conversion efficiency with temperature. For temperatures corresponding to the hot, demagnetized regime of interaction, the magnetic energy shows low amplification of the order of 5 per cent of the initial drift kinetic energy. 

\subsection{Wave-particles interactions}
\label{sec:wave_interaction}
\begin{figure}
\includegraphics[width=\columnwidth]{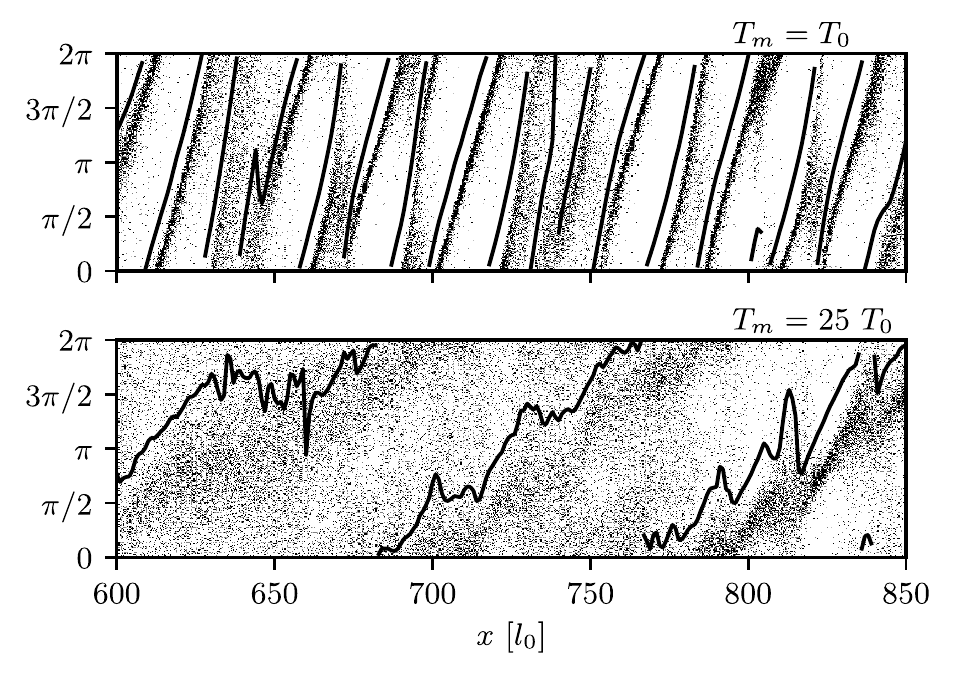}
\caption{Perturbed magnetic field phase $\phi_B=\tan^{-1}(B_z/B_y)$ (black solid line) and main protons particle velocity phase $\phi_v=\tan^{-1}(v_z/v_y)$ (black dots) as a function of space (from  $600$ to $800\ x/l_0$) for 1D simulations, during the linear growth phase phase. Upper panel: $T_m=T_0$, lower panel: $T_m=25\ T_0$.}  
\label{fig:figure_6}
\end{figure}

The instability relies on the helicity of the perturbed magnetic field and background fluid velocity field. Both interact via the induced electric field and generate a positive feedback with one another, destabilizing the electromagnetic wave. The resulting main protons velocity field is a helix with negative helicity, left-hand polarization, positive direction of propagation and with a norm increasing exponentially over time, whereas the perturbed magnetic field forms expanding field lines with negative helicity, right-hand polarization and negative direction of propagation (see appendix \ref{sec:appendix_hel}). These interlaced structures can be observed in Fig. \ref{fig:figure_6}, where the main protons particle velocity phase $\phi_v=\tan^{-1}(v_z/v_y)$ is plotted along with the perturbed magnetic field phase $\phi_B=\tan^{-1}(B_z/B_y)$ at the beginning of the non-linear growth phase, for low temperature $T_m=T_0$ and high temperature $T_m=25\ T_0$. 

The positive slope of the perturbed magnetic field phase $\phi_B$ illustrates its negative helicity in the ($\vec e_y,\vec e_z$) plane. In the cold case (Fig. \ref{fig:figure_6} upper panel) it is clear that the velocity phase $\phi_v$ also develops a coherent structure that closely follows that of the magnetic field. At a given position $x$, the velocity phase of all the particles bunches around a well defined value, for example at $x = 625l_0$ the velocity phase is $\phi_v \approx \pi$. The phase shift between the magnetic field and the velocity field at a given position, $\Delta\phi=\phi_B-\phi_v$, is close to $\pi/2$, which was expected from Eq. \ref{eq:momentum_heur}. Indeed the cosmic rays driving term $-\vec j_{cr}\times\vec B_1/\rho$ accelerates the background fluid in a direction perpendicular to both the ambient and local perturbed magnetic field. The phase shift $\Delta\phi$ is observed to remain constant during the linear evolution of the instability, corresponding to a coherent motion of the main protons with respect to the electromagnetic wave.

Increasing the temperature does not modify the helicity of the perturbation (Fig. \ref{fig:figure_6} lower panel), and we see an increase in wavelength as predicted by linear theory. We also find a less clearly defined phase bunching of the main protons velocity compared to the cold regime. The main protons high mobility in the demagnetized regime allows them to migrate quickly along the ambient magnetic field, mixing up the phase shift between the background velocity and the magnetic field fluctuations. As a result, the induced electric field is weakened and the instability grows less efficiently. This may constitute a possible physical interpretation to the monotonously decreasing growth rate with thermal velocity found in the hot regime (Eq. \ref{eq:gamma_hot_analytic}). 

\begin{figure}
	\includegraphics[width=\columnwidth]{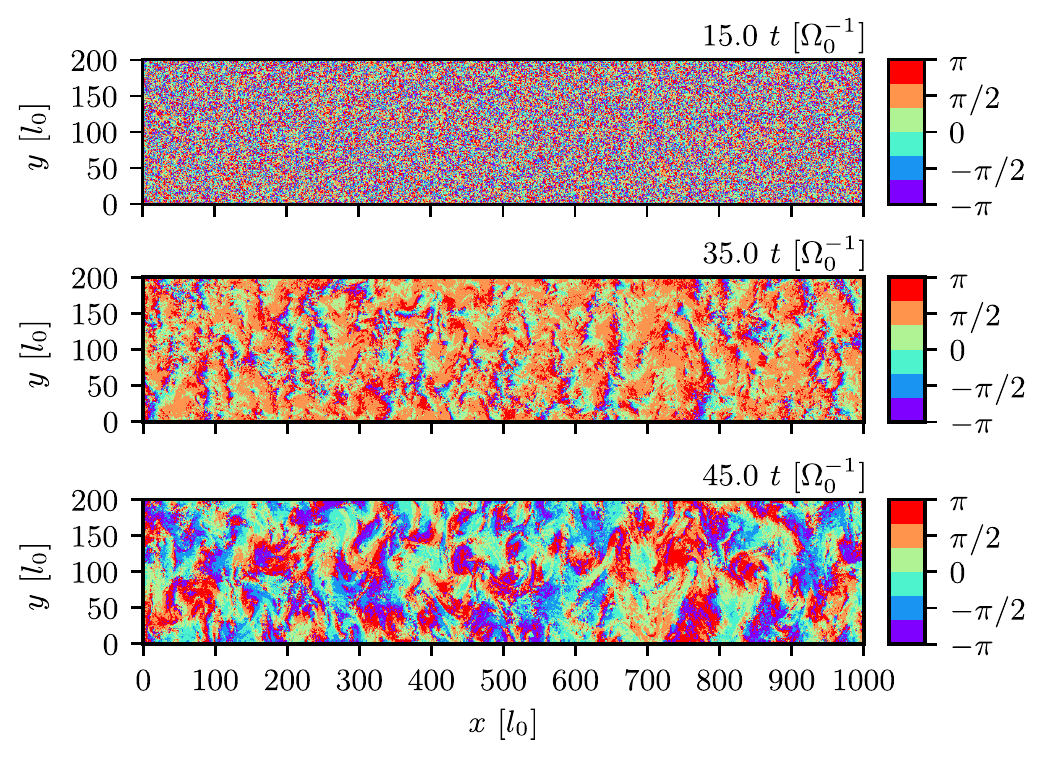}
    \caption{Magnetic field and main protons velocity phase difference $\Delta\phi$ map at three different times: beginning of the linear growth phase ($t=15\ \Omega_0^{-1}$), during the linear growth phase ($t=35\ \Omega_0^{-1}$), after saturation ($t=45\ \Omega_0^{-1}$). The difference is calculated locally as $\Delta\phi=\tan^{-1}(\sin(\phi_B-\phi_v)/\cos(\phi_B-\phi_v))$. Obtained from a 2D simulations with $T_m=25\ T_0$. The theoretical prediction from the fluid model yields a uniform phase shift $\Delta\phi=\pi/2$ and we recover this result in the simulations, with modulations due to the high temperature of the main protons.}
    \label{fig:figure_7}
\end{figure}

These effects can also be observed and quantified in 2D simulations. Fig. \ref{fig:figure_7} presents the phase difference between the fluid velocity phase and magnetic field phase $\Delta\phi=\phi_B-\phi_v$ for three different times: $t=15\ \Omega_0^{-1}$, $t=35\ \Omega_0^{-1}$ and $t=45\ \Omega_0^{-1}$, corresponding to the beginning and end of the linear growth phase, and after magnetic saturation. The main protons temperature is $T_m=25\ T_0$. Other parameters are described in Table \ref{table:parameters_simulations}. During the early times of growth (upper panel), the magnetic field and background velocity field are essentially uncorrelated and the growth is slow. By the end of the linear phase (middle panel), the phase shifttends to the expected value $\Delta\phi=\phi_B-\phi_v=\pi/2$ and allows the fast growth of the perturbation. After the magnetic field saturation (lower panel), it randomizes as the cosmic rays magnetic force no longer imposes the magnetic field rotation to the background fluid.

\begin{figure}
	\includegraphics[width=\columnwidth]{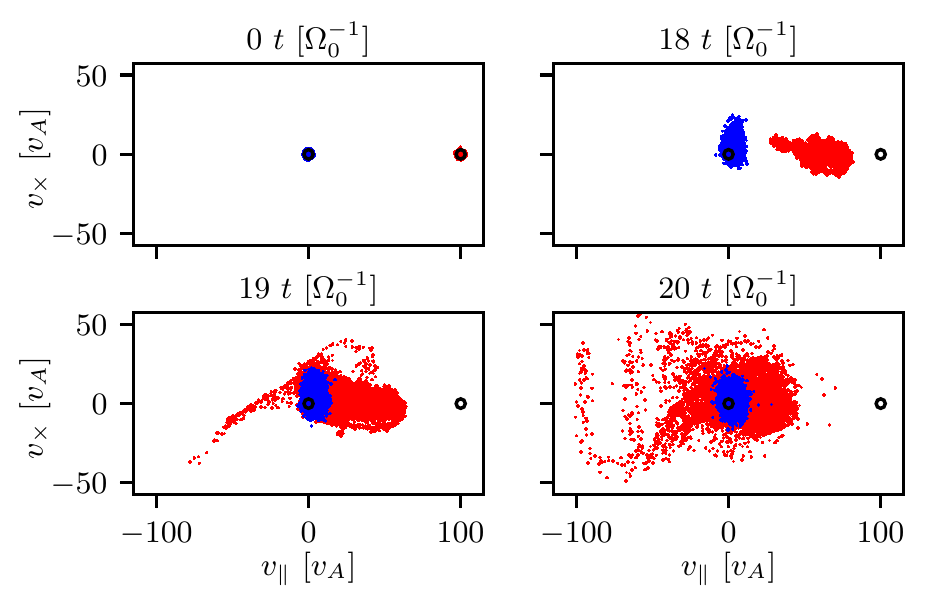}
    \caption{Distribution of the main protons (blue dots) and cosmic rays (red dots) in $(v_\parallel,v_\times)$ space, during the instability non-linear phase. Obtained from a 1D simulation with a main protons temperature $T_m=T_0$. The black circles indicate the initial velocity dispersion.}
    \label{fig:figure_8}
\end{figure}

The instability leads to the development of important anisotropies in the protons velocity distributions for both the main and the cosmic rays populations which cannot be described in isotropic fluid simulations. In the following, we will make use of the magnetic field aligned basis $(\vec e_\parallel,\vec e_\times,\vec e_\perp)$ presented in the previous section to describe these effects. Fig. \ref{fig:figure_8} shows the distributions of the proton populations in $(v_\parallel,v_\times)$ space during the non-linear phase of the instability growth. The parallel electric field induced by the main protons rotating motion $\vec E_{\parallel}=-\vec u_1\times \vec B_1$ (Eq. \ref{eq:induction}) decelerates the cosmic rays (upper right panel), leading to  particles acquiring a velocity in the opposite direction to their original one (in the reference frame of the initially immobile main protons, lower left panel). The cosmic rays are then able to interact resonantly as the resonance condition $\omega_r-ku_{\parallel\alpha}+p^\pm\Omega_{\alpha}=0$ is fulfilled with the right-hand polarized backward propagating waves. As a result, the cosmic rays are strongly scattered in the ($\vec e_y,\vec e_z$) plane (lower right panel). This effect is highly non-linear: the cosmic rays destabilize electromagnetic waves in a non-resonant way, and interact later on with the large amplitude waves they have generated. The main protons acceleration in the normal direction $\vec e_{\times}$ (upper right panel) is well observed, and is correlated to the slow down of the cosmic rays. We note that the cosmic rays velocity distribution initially Maxwellian is greatly altered during the linear and non-linear evolution, and returns to equilibrium only during the relaxation phase. 

\begin{figure}
	\includegraphics[width=\columnwidth]{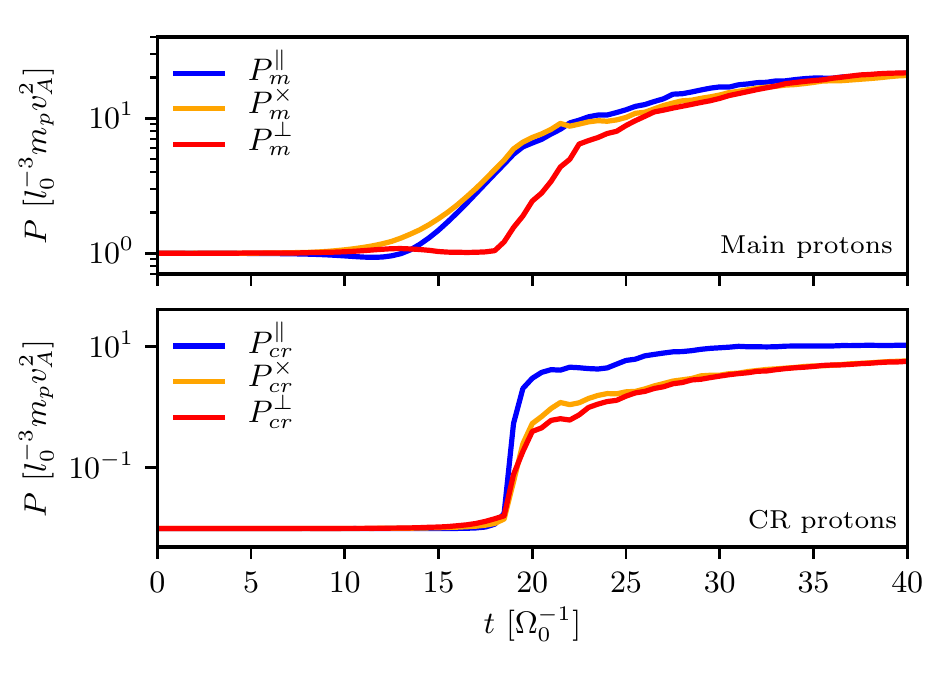}
    \caption{Diagonal terms of the pressure tensor in the local magnetic field aligned basis for the main protons (upper panel) and cosmic rays (lower panel), as a function of time between $t=0\ \Omega_0^{-1}$ and $t=40\ \Omega_0^{-1}$, for a main protons temperature $T_m=T_0$. The blue, orange and red curves corresponds to the parallel $P^\parallel$, normal $P^\times$ and perpendicular $P^\perp$ components respectively. The values are calculated locally, then averaged over the simulation domain. Obtained from a 1D simulation with a main protons temperature $T_m=T_0$.}
    \label{fig:figure_9}
\end{figure}

The heating and scattering can be quantified by investigating the time evolution of the diagonal terms $ii$ of the protons pressure tensor over time, defined as:
\begin{align}
P_{ii,\ell}=m_\ell W_\ell \sum_kS(\vec{x}-\vec{x}_k)(v_{i\ell,k}-u_{i\ell})^2
\label{eq:pressure}
\end{align}
where $m_\ell$ and $W_\ell$ are the mass and numerical weight of the proton population $\ell$, and $S(\vec{x}-\vec{x}_k)$ is the first order B-spline. The sum is calculated over all the macro-particles $k$ of population $\ell$. Fig. \ref{fig:figure_9} shows the diagonal components in the magnetic field aligned basis for the main protons (upper panel) and the cosmic rays (lower panel) in the low temperature case $T_m=T_0$. The main protons pressure starts increasing in the parallel and normal direction first, as magnetic perturbations become of the same order as the initial magnetic field. The perpendicular direction is not heated. This non-gyrotropic behaviour is a product of the electric field fluctuations, which are oriented along the normal direction because of Faraday's law and generate heating in this direction. As the main protons rotate with the electromagnetic perturbation, the perpendicular direction sees no electric fluctuations and remains unheated. We obtain the same pressure anisotropies in 2D simulations (not shown here). The pressure gradient generated between the normal and perpendicular components may act against the main protons rotation, thus reducing the growth rate of the instability as well as the saturation level. As a consequence, introducing collisions between main protons at frequencies comparable to the NR mode growth rate might favour the growth of the NR mode, by isotropizing the pressure components and reducing the pressure gradient counter force. 

The beginning of the collision-less isotropization process (around 18 $\Omega_0^{-1}$) occurs at the transition from linear to non-linear growth phase, when the main protons rotating motion starts lagging behind with the electromagnetic waves. The increase in perpendicular pressure is a consequence of the rotation of the magnetic field aligned basis as it starts being uncorrelated to the main protons motion. The pressure observed in the normal direction is projected along the perpendicular component, hence the increase in the perpendicular component and the reduced growth of the normal one. Pressure anisotropies are suppressed when the phase shift $\Delta\phi=\phi_B-\phi_v$ is totally randomized during the non-linear and relaxation phases. The main protons are then slowly heated during the relaxation phase. By supposing a perfect gas behaviour, we can estimate the temperature as $k_bT_m = P_m/n_m$, and obtain values corresponding to one order of magnitude increase with respect to the initial one. Simulations with larger initial main protons temperatures show less heating as the instability develops less efficiently.

We observe that the cosmic rays pressure increase in the $(\hat{e}_\times,\hat{e}_\perp)$ plane takes place when a fraction of the streaming particles acquire a negative velocity. This corresponds to the sudden increase in pressure, in agreement with the previous discussion. The cosmic rays velocity dispersion during the linear growth phase seen in $(v_\parallel,v_\times)$ space (Fig. \ref{fig:figure_8} upper right panel) is an effect of the velocity space representation, which does not take into account the position of the macro-particles. The cosmic rays follow an organized motion as no pressure increase occurs, becoming stochastic after resonant interactions begin settling in during the non-linear phase (Fig. \ref{fig:figure_8} lower left panel).

\begin{figure}
    \includegraphics[width=\columnwidth]{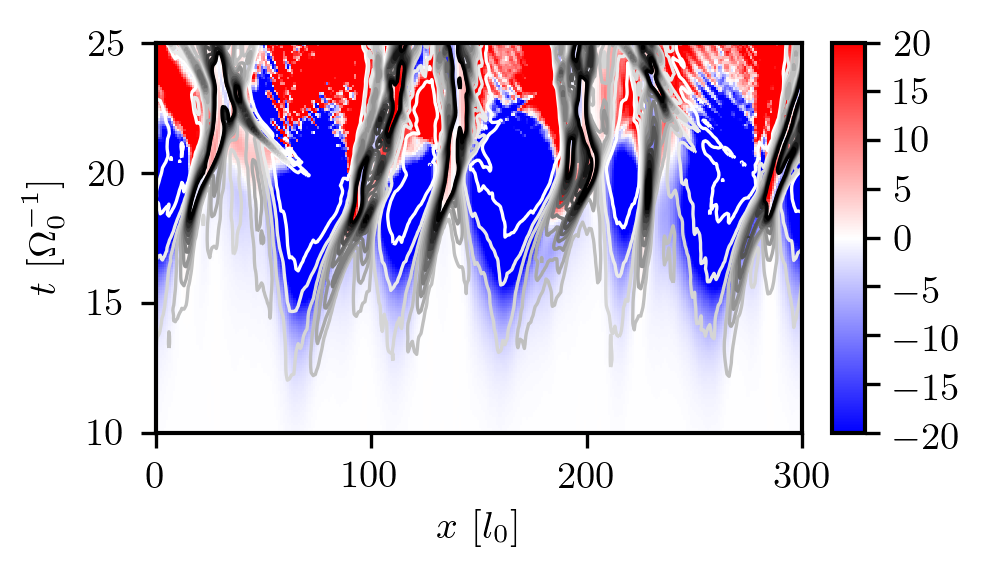}
    \caption{Parallel electric field component $E_{\parallel}$ (in unit of $v_{A0}B_0$, blue to red color scale) and main protons density $n_m$ (white $n_m=0.5\ n_0$ to black $n_m=3\ n_0$ contours) as a function of space (from 0 to 300 $x/l_0$, abscissa) and time (from 10 to 25 $t\Omega_0$, ordinate), for a 1D simulation with a main protons temperature $T_m=T_0$. The magnetic field saturation is reached at $t=21\ \Omega_0^{-1}$.}
    \label{fig:figure_10}
\end{figure}

The heating in the parallel direction has a different origin, and can be linked to the electric field induced by the rotating motion of the main protons. A study of the relative intensities of the longitudinal $\vec k\cdot\vec E/k$ and transverse $\vec k\times\vec E/k$ electric field reveals an important electrostatic component developing during the non-linear phase. The longitudinal electric field (corresponding to the parallel component in 1D simulations) spatial and temporal evolution is presented in Fig. \ref{fig:figure_10} alongside with the main protons density. Regions of fast growing mode appear well delimited in space, and expand in both negative and positive directions. Large electric field gradients are generated, leading to an important heating of both protons populations. The background fluid is accelerated in the same direction as the cosmic rays initial velocity as it is negatively charged. Because of the continuity equation, the background plasma accumulates mass on the right of the growing electric field regions. As a consequence, large density fluctuations are generated with cavities of low density, correlated with regions of fast growing modes and important heating of both protons populations. Note the reversal of the electric field after saturation ($t_{sat}=21\ \Omega_0^{-1}$), corresponding to the main protons normal velocity $u_m^\times$ changing sign in the magnetic field aligned basis (Fig. \ref{fig:figure_4}, lower panel) and inducing a positive electric field. As a consequence the cosmic rays drift kinetic energy cannot be converted into magnetic energy, leading to the NR mode saturation as discussed previously.

\begin{figure*}
\includegraphics[width=\textwidth]{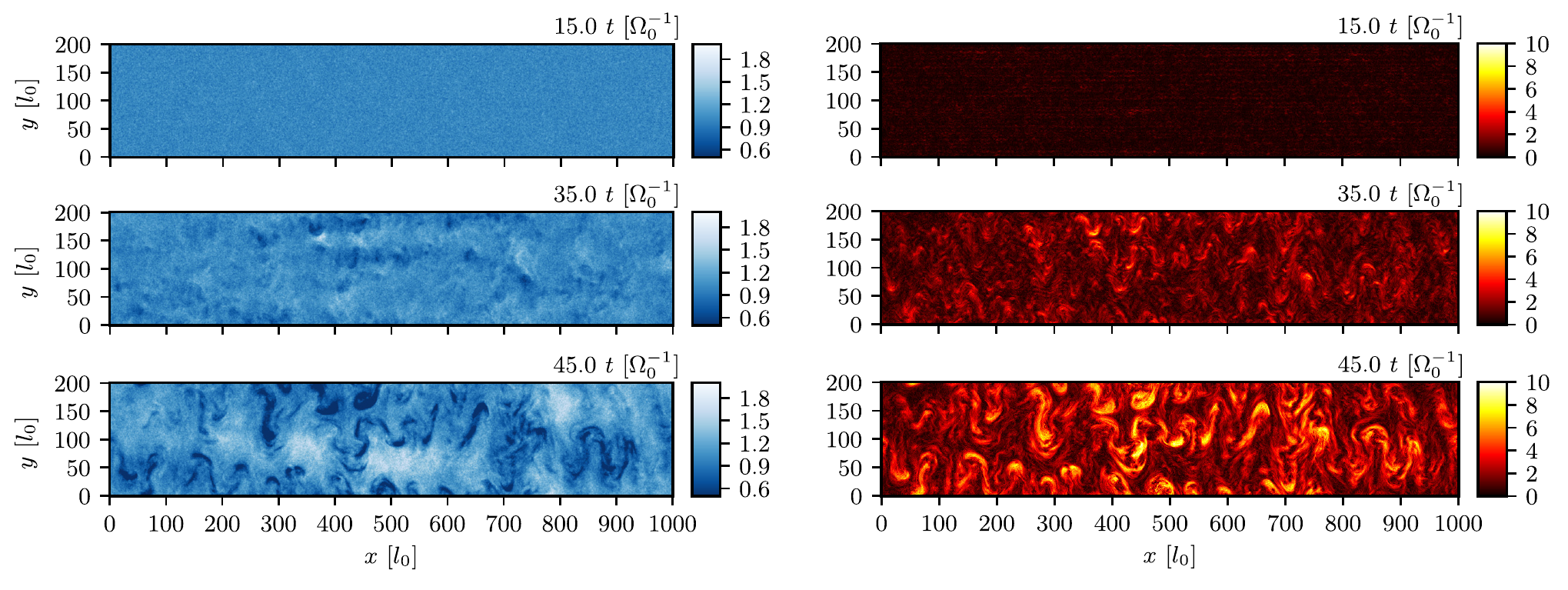}
\caption{Main protons density $n_m/n_0$ map (left panels) and perturbed magnetic field intensity $B_1=||\vec{B}-\vec{B}_0||/B_0$ map (right panels) at three different times: beginning of the linear growth phase ($t=15\ \Omega_0^{-1}$), during the linear growth phase ($t=35\ \Omega_0^{-1}$), after saturation ($t=45\ \Omega_0^{-1}$). Obtained from a 2D simulation with $T_m=25\ T_0$.}
\label{fig:figure_11}
\end{figure*}

The 2D simulations bring additional information on the main protons density spatial structures. Our simulations results are presented in Fig. \ref{fig:figure_11}. Density fluctuations are found to increase in scale from tenth to hundredth of $l_0$ over time, as small scale density holes along the initial magnetic field direction (observed in 1D simulations) merge together to generate large scale fluctuations during the non-linear evolution of the instability. The density holes expand in the perpendicular plane because of the increasing magnetic pressure, generating density fluctuations up to $n_m/n_0\sim 2$ in the background plasma at the same spatial scales as the magnetic fluctuations, on the order of a hundredth of the proton inertial length for the parameters investigated. This result agrees with previous studies using a fluid description (\citealt{bell_cosmic_2013}, \citealt{bai_magnetohydrodynamic-particle--cell_2015}), and may play a role in allowing further magnetic field amplification, by taking into account potentially important dynamo effects at supernova shocks (\citealt{del_valle_turbulence-induced_2016}).

\section{Summary and discussion}
\label{sec:conclusion}
The non-resonant cosmic rays streaming instability has drawn much attention as an efficient sources of large amplitude magnetic field fluctuations. Its presence is thought to be ubiquitous in many space and astrophysical environments with hugely varying physical conditions (temperature, magnetic field, etc). The basic mechanism of the instability may be simply captured within a fluid model consisting of a negatively charged background plasma immersed in a large scale magnetic field. The fluid supports an electric current that acts to compensate the current generated by the streaming cosmic rays and which ultimately drives the instability. However a fluid model for the background plasma neglects kinetic effects which may be crucial to correctly describe both the linear and non-linear evolution of the NR mode. In this work we have explored the effects of plasma temperature on the development of the NR mode. 

Within the framework of linear kinetic theory, we have extended the existing theory of the instability from zero or small main protons temperatures to the hot, demagnetized regime of interaction, and proposed analytical expressions for the growth rate and associated angular frequency, and wave number. In particular, we found that the temperature dependence of the growth rate of fastest growing mode changes from $T^{-1/3}$ for relatively small temperatures (warm regime), to a steeper $T^{-1/2}$ at higher temperatures (hot, demagnetized regime). The threshold for the hot, demagnetized regime is estimated from the Larmor radius of the main protons and the maximum unstable wave number as $k_{\ms cold}r_{Lm}>2$, which can be rewritten as $\beta_m/2>(4n_mv_{A0}/n_{cr}u_{\parallel cr})^2$ where we have defined the plasma $\beta_m=2(v_{Tm}/v_{A0})^2$. 

Using a density ratio $n_{cr}/n_m=10^{-5}$ and a shock velocity $u_{\parallel cr}=2.10^{3}$ km.s$^{-1}$ typically considered in supernova and galaxy clusters shocks, one immediately finds that very large plasma $\beta_m$ are required to reach the demagnetized regime. For typical parameters of the interstellar medium, $n_m=1$ cm$^{-3}$, $T_m=10^4$ K and $B=10^{-6}$ G the demagnetized regime is not relevant even by considering locally smaller magnetic field, and larger temperatures such as those found in superbubbles (\citealt{mac_low_superbubbles_1988}). The picture may change however when considering leakage of cosmic rays in the intergalactic medium. Taking parameters $n_m=10^{-6}$ cm$^{-3}$, $n_{cr}=10^{-9}$ cm$^{-3}$, $T_m=10^6$ K, $u_{\parallel cr}=10^2$ km.s$^{-1}$, and a magnetic field $B=10^{-11}$ G (\citealt{kulsrud_origin_2008}), one obtains a plasma $\beta_m$ larger than the required value for the main protons to be demagnetized, and finds a growth rate $\gamma_{hot}=6.6\times10^{-10}$ s$^{-1}$ from Eq. \ref{eq:gamma_hot_analytic} corresponding to a growth time of the order of $2\pi\gamma_{hot}^{-1}=300$ years. The growth rate is strongly reduced by temperature by a factor $(2\pi)^{1/2}v_{A0}/v_{Tm}=3\times 10^{-3}$ compared to the cold prediction, but still larger than the cosmic rays advection rate (\citealt{zweibel_environments_2010}), which implies that the non-resonant mode could develop in such a medium despite the background protons being demagnetized. 

To explore the non-linear evolution of the instability, we have performed 1D and 2D hybrid-PIC simulations for main protons temperatures spanning over three orders of magnitude, allowing us to probe the cold, warm, and hot regimes of interaction. Both analytical and numerical results show that because of the finite main protons Larmor radius, the unstable waves at small scales are damped with increasing temperature. This leads to a shift of the unstable wavelengths toward larger scales and to an overall slower growth of the magnetic field perturbations. Another important kinetic effect, which to our knowledge was not discussed in the literature before, is the development of an important anisotropic heating of the background protons during the linear phase of growth. It occurs between the two components in the plane perpendicular to the ambient magnetic field, and is generated by the distinctive geometrical correlation between the electric and background fluid velocity fields during the growth of the instability. This result suggests that MHD and MHD-PIC simulations with isotropic closure may not be adequate to describe all of the essential features of the NR mode, as a more sophisticated closure may be required to reproduce the anisotropic heating.

We have highlighted the existence of a non-linear phase of amplification of the magnetic field which follows the growth predicted by linear theory. This phase starts once the $k_{\min}=k_{\max}$ condition is fulfilled, and ends with the loss of correlation between the magnetic perturbation and the background fluid motion, which halts the growth of the instability. This saturation mechanism is well observed in all the simulations. A quantitative prediction of the saturated magnetic field intensity remains a challenging issue. An estimate can be obtained by extrapolating quasi-linear theory results (\citealt{winske_diffuse_1984}, \citealt{winske_electromagnetic_1986}) to saturation, which yields that half of the initial cosmic rays drift kinetic energy should be converted into magnetic field energy (Eq. \ref{eq:saturation}). We find in the simulations that the energy conversion efficiency between the streaming population drift kinetic energy and the magnetic field energy can be up to $60$ per cent in the low temperature case, corresponding to one order of magnitude increase in comparison to the initial ambient magnetic field with our parameters. This is strongly reduced with increasing temperature, down to less than $5$ per cent in the hot and demagnetized regime.

The non-resonant mode growth leads to the generation of large density gradients in the background plasma during the non-linear phase, produced by the induced parallel electric field and by the increasing magnetic pressure. In addition, the cosmic rays are decelerated by this parallel electric field and some of them are then able to interact with the large amplitude electromagnetic waves they have generated, leading to important scattering in velocity space. As such, the non-resonant mode contains an intrinsic scattering mechanism, which may play a role in the efficient confinement of the cosmic rays at the shock boundary of supernova remnants. 

We conclude by noting that as in many previous studies (e.g. \citealt{winske_diffuse_1984}, \citealt{riquelme_non-linear_2009}, \citealt{ohira_two-dimensional_2009}), we performed simulations without a continuous injection of streaming particles. The immediate consequence is that the cosmic rays current is self-consistently decreasing through time as the drift kinetic energy is being converted into magnetic fluctuations. An alternative approach is to maintain the driving current, either by re-accelerating the cosmic rays artificially (\citealt{lucek_non-linear_2000}), or by injecting new ones in the simulation domain over time (\citealt{bai_magnetohydrodynamic-particle--cell_2015}, \citealt{mignone_particle_2018}, \citealt{casse_magnetic_2018}) which was used to directly simulate particles acceleration at supernova shocks. A comparison between these approaches shows that the development of the NR instability is not significantly altered. In particular the magnetic field intensity at saturation and the density fluctuations are quantitatively similar, with magnetic field amplifications of the order of ten times the ambient magnetic field and large density fluctuation of the order of the initial plasma density. These results however apply to the cold regime, and the ambient medium temperature remains an important factor in determining whether the non-resonant streaming instability can efficiently generate magnetic field fluctuations, and should be taken into account to model accurately cosmic rays acceleration with realistic plasma conditions.

\section*{Acknowledgements}
We thank Elena Amato for her hospitality at the Arcetri Observatory and, together with Stefano Gabici, for useful discussions. This work was performed using HPC resources from GENCI- [TGCC] (Grant 2019- [DARI A0060410819]). This work was granted access to the HPC resources of MesoPSL financed by the Region Ile de France and the project EquipMeso (reference ANR-10-EQPX29-01) of the programme Investissements d'Avenir supervised by the Agence Nationale pour la Recherche. This work was partly done within the Plas@Par LABEX project and supported by grant 11-IDEX-0004-02 from ANR (France).

\section*{Data availability}
The data underlying this article will be shared on reasonable request to the corresponding author.

\bibliographystyle{mnras}
\bibliography{references} 

\begin{thebibliography}{}
\makeatletter
\relax
\def\mn@urlcharsother{\let\do\@makeother \do\$\do\&\do\#\do\^\do\_\do\%\do\~}
\def\mn@doi{\begingroup\mn@urlcharsother \@ifnextchar [ {\mn@doi@}
  {\mn@doi@[]}}
\def\mn@doi@[#1]#2{\def\@tempa{#1}\ifx\@tempa\@empty \href
  {http://dx.doi.org/#2} {doi:#2}\else \href {http://dx.doi.org/#2} {#1}\fi
  \endgroup}
\def\mn@eprint#1#2{\mn@eprint@#1:#2::\@nil}
\def\mn@eprint@arXiv#1{\href {http://arxiv.org/abs/#1} {{\tt arXiv:#1}}}
\def\mn@eprint@dblp#1{\href {http://dblp.uni-trier.de/rec/bibtex/#1.xml}
  {dblp:#1}}
\def\mn@eprint@#1:#2:#3:#4\@nil{\def\@tempa {#1}\def\@tempb {#2}\def\@tempc
  {#3}\ifx \@tempc \@empty \let \@tempc \@tempb \let \@tempb \@tempa \fi \ifx
  \@tempb \@empty \def\@tempb {arXiv}\fi \@ifundefined
  {mn@eprint@\@tempb}{\@tempb:\@tempc}{\expandafter \expandafter \csname
  mn@eprint@\@tempb\endcsname \expandafter{\@tempc}}}

\bibitem[\protect\citeauthoryear{Akimoto, Winske, Gary  \& Thomsen}{Akimoto
  et~al.}{1993}]{akimoto_nonlinear_1993}
Akimoto K.,  Winske D.,  Gary S.~P.,   Thomsen M.~F.,  1993, \mn@doi [Journal
  of Geophysical Research: Space Physics] {10.1029/92JA02345}, 98, 1419

\bibitem[\protect\citeauthoryear{Amato \& Blasi}{Amato \&
  Blasi}{2009}]{amato_kinetic_2009}
Amato E.,  Blasi P.,  2009, \mn@doi [Monthly Notices of the Royal Astronomical
  Society] {10.1111/j.1365-2966.2008.14200.x}, 392, 1591

\bibitem[\protect\citeauthoryear{Bai, Caprioli, Sironi  \& Spitkovsky}{Bai
  et~al.}{2015}]{bai_magnetohydrodynamic-particle--cell_2015}
Bai X.-N.,  Caprioli D.,  Sironi L.,   Spitkovsky A.,  2015, \mn@doi [The
  Astrophysical Journal] {10.1088/0004-637X/809/1/55}, 809, 55

\bibitem[\protect\citeauthoryear{Bell}{Bell}{2004}]{bell_turbulent_2004}
Bell A.~R.,  2004, \mn@doi [Monthly Notices of the Royal Astronomical Society]
  {10.1111/j.1365-2966.2004.08097.x}, 353, 550

\bibitem[\protect\citeauthoryear{Bell}{Bell}{2013}]{bell_cosmic_2013}
Bell A.~R.,  2013, \mn@doi [Astroparticle Physics]
  {10.1016/j.astropartphys.2012.05.022}, 43, 56

\bibitem[\protect\citeauthoryear{Boris}{Boris}{1970}]{boris_acceleration_1970}
Boris J.~P.,  1970, Proceedings of the fourth conference on numerical
  simulation of plasmas

\bibitem[\protect\citeauthoryear{Casse, van Marle  \& Marcowith}{Casse
  et~al.}{2018}]{casse_magnetic_2018}
Casse F.,  van Marle A.~J.,   Marcowith A.,  2018, \mn@doi [Plasma Physics and
  Controlled Fusion] {10.1088/1361-6587/aa8482}, 60, 014017

\bibitem[\protect\citeauthoryear{Crumley, Caprioli, Markoff  \&
  Spitkovsky}{Crumley et~al.}{2019}]{crumley_kinetic_2019}
Crumley P.,  Caprioli D.,  Markoff S.,   Spitkovsky A.,  2019, \mn@doi [Monthly
  Notices of the Royal Astronomical Society] {10.1093/mnras/stz232}, 485, 5105

\bibitem[\protect\citeauthoryear{Fried \& Conte}{Fried \&
  Conte}{1961}]{fried_plasma_1961}
Fried B.~D.,  Conte S.~D.,  1961, The {Plasma} {Dispersion} {Function}.
Elsevier

\bibitem[\protect\citeauthoryear{Gary \& Feldman}{Gary \&
  Feldman}{1978}]{gary_second-order_1978}
Gary S.~P.,  Feldman W.~C.,  1978, \mn@doi [Physics of Fluids]
  {10.1063/1.862081}, 21, 72

\bibitem[\protect\citeauthoryear{Gary, Smith, Lee, Goldstein  \& Forslund}{Gary
  et~al.}{1984}]{gary_electromagnetic_1984}
Gary S.~P.,  Smith C.~W.,  Lee M.~A.,  Goldstein M.~L.,   Forslund D.~W.,
  1984, \mn@doi [The Physics of Fluids] {10.1063/1.864797}, 27, 1852

\bibitem[\protect\citeauthoryear{Kulsrud \& Pearce}{Kulsrud \&
  Pearce}{1969}]{kulsrud_effect_1969}
Kulsrud R.,  Pearce W.~P.,  1969, \mn@doi [The Astrophysical Journal]
  {10.1086/149981}, 156, 445

\bibitem[\protect\citeauthoryear{Kulsrud \& Zweibel}{Kulsrud \&
  Zweibel}{2008}]{kulsrud_origin_2008}
Kulsrud R.~M.,  Zweibel E.~G.,  2008, \mn@doi [Reports on Progress in Physics]
  {10.1088/0034-4885/71/4/046901}, 71, 046901

\bibitem[\protect\citeauthoryear{Lucek \& Bell}{Lucek \&
  Bell}{2000}]{lucek_non-linear_2000}
Lucek S.~G.,  Bell A.~R.,  2000, \mn@doi [Monthly Notices of the Royal
  Astronomical Society] {10.1046/j.1365-8711.2000.03363.x}, 314, 65

\bibitem[\protect\citeauthoryear{Mac~Low \& McCray}{Mac~Low \&
  McCray}{1988}]{mac_low_superbubbles_1988}
Mac~Low M.-M.,  McCray R.,  1988, \mn@doi [The Astrophysical Journal]
  {10.1086/165936}, 324, 776

\bibitem[\protect\citeauthoryear{Matthews, Bell, Blundell  \& Araudo}{Matthews
  et~al.}{2017}]{matthews_amplification_2017}
Matthews J.~H.,  Bell A.~R.,  Blundell K.~M.,   Araudo A.~T.,  2017, \mn@doi
  [Monthly Notices of the Royal Astronomical Society] {10.1093/mnras/stx905},
  469, 1849

\bibitem[\protect\citeauthoryear{Mignone, Bodo, Vaidya  \& Mattia}{Mignone
  et~al.}{2018}]{mignone_particle_2018}
Mignone A.,  Bodo G.,  Vaidya B.,   Mattia G.,  2018, \mn@doi [The
  Astrophysical Journal] {10.3847/1538-4357/aabccd}, 859, 13

\bibitem[\protect\citeauthoryear{Ohira, Reville, Kirk  \& Takahara}{Ohira
  et~al.}{2009}]{ohira_two-dimensional_2009}
Ohira Y.,  Reville B.,  Kirk J.~G.,   Takahara F.,  2009, \mn@doi [The
  Astrophysical Journal] {10.1088/0004-637X/698/1/445}, 698, 445

\bibitem[\protect\citeauthoryear{Onsager, Winske  \& Thomsen}{Onsager
  et~al.}{1991}]{onsager_interaction_1991}
Onsager T.~G.,  Winske D.,   Thomsen M.~F.,  1991, \mn@doi [Journal of
  Geophysical Research: Space Physics] {10.1029/90JA02008}, 96, 1775

\bibitem[\protect\citeauthoryear{Pelletier, Lemoine  \& Marcowith}{Pelletier
  et~al.}{2006}]{pelletier_turbulence_2006}
Pelletier G.,  Lemoine M.,   Marcowith A.,  2006, \mn@doi [Astronomy \&
  Astrophysics] {10.1051/0004-6361:20054737}, 453, 181

\bibitem[\protect\citeauthoryear{Reville, Kirk, Duffy  \& O'Sullivan}{Reville
  et~al.}{2008}]{reville_environmental_2008}
Reville B.,  Kirk J.~G.,  Duffy P.,   O'Sullivan S.,  2008, \mn@doi
  [International Journal of Modern Physics D] {10.1142/S021827180801342X}, 17,
  1795

\bibitem[\protect\citeauthoryear{Riquelme \& Spitkovsky}{Riquelme \&
  Spitkovsky}{2009}]{riquelme_non-linear_2009}
Riquelme M.~A.,  Spitkovsky A.,  2009, \mn@doi [The Astrophysical Journal]
  {10.1088/0004-637X/694/1/626}, 694, 626

\bibitem[\protect\citeauthoryear{Scharer}{Scharer}{1967}]{scharer_cyclotron_1967}
Scharer J.~E.,  1967, \mn@doi [Physics of Fluids] {10.1063/1.1762153}, 10, 591

\bibitem[\protect\citeauthoryear{Sentman, Edmiston  \& Frank}{Sentman
  et~al.}{1981}]{sentman_instabilities_1981}
Sentman D.~D.,  Edmiston J.~P.,   Frank L.~A.,  1981, \mn@doi [Journal of
  Geophysical Research: Space Physics] {10.1029/JA086iA09p07487}, 86, 7487

\bibitem[\protect\citeauthoryear{{Smets, R.}}{{Smets,
  R.}}{2020}]{smets_r_heckle_2020}
{Smets, R.} 2020, Heckle.
GitHub, \url {https://github.com/rochSmets/heckle}

\bibitem[\protect\citeauthoryear{Winske \& Leroy}{Winske \&
  Leroy}{1984}]{winske_diffuse_1984}
Winske D.,  Leroy M.~M.,  1984, \mn@doi [Journal of Geophysical Research: Space
  Physics] {10.1029/JA089iA05p02673}, 89, 2673

\bibitem[\protect\citeauthoryear{Winske \& Quest}{Winske \&
  Quest}{1986}]{winske_electromagnetic_1986}
Winske D.,  Quest K.~B.,  1986, \mn@doi [Journal of Geophysical Research: Space
  Physics] {10.1029/JA091iA08p08789}, 91, 8789

\bibitem[\protect\citeauthoryear{Zacharegkas, Caprioli  \&
  Haggerty}{Zacharegkas et~al.}{2019}]{zacharegkas_modeling_2019}
Zacharegkas G.,  Caprioli D.,   Haggerty C.,  2019, arXiv:1909.06481 [astro-ph,
  physics:physics]

\bibitem[\protect\citeauthoryear{Zirakashvili, Ptuskin  \& Voelk}{Zirakashvili
  et~al.}{2008}]{zirakashvili_modeling_2008}
Zirakashvili V.~N.,  Ptuskin V.~S.,   Voelk H.~J.,  2008, \mn@doi [The
  Astrophysical Journal] {10.1086/529579}, 678, 255

\bibitem[\protect\citeauthoryear{Zweibel \& Everett}{Zweibel \&
  Everett}{2010}]{zweibel_environments_2010}
Zweibel E.~G.,  Everett J.~E.,  2010, \mn@doi [The Astrophysical Journal]
  {10.1088/0004-637X/709/2/1412}, 709, 1412

\bibitem[\protect\citeauthoryear{del Valle, Lazarian  \& Santos-Lima}{del Valle
  et~al.}{2016}]{del_valle_turbulence-induced_2016}
del Valle M.~V.,  Lazarian A.,   Santos-Lima R.,  2016, \mn@doi [Monthly
  Notices of the Royal Astronomical Society] {10.1093/mnras/stw340}, 458, 1645

\makeatother
\end{thebibliography}





\appendix

\onecolumn

\begin{center}
\section{Helicity and polarization}
\label{sec:appendix_hel}
\end{center}
We consider an electromagnetic, circularly polarized perturbation propagating along the $\vec e_x$ direction with an angular frequency $\omega=\omega_r+i\gamma$ where $\omega_r$ is defined to be positive. The wave number $k$ can be either positive or negative depending on the direction of propagation. We define the polarization as the sense or rotation of the magnetic field in time, observed at a given position in space such that $\vec B_1=B_1\cos(kx-p^{\pm}\omega t)\vec e_y -B_1\sin(kx-p^{\pm}\omega t)\vec e_z$, where $p^\pm$ corresponds to the polarization of the wave: $p^\pm=+1$ for a right hand polarized wave and $p^\pm=-1$ for a left hand polarized wave.  We define the helicity of a wave, as the sense of rotation of the magnetic field in space, at a given time. Helicity and polarization are linked through the direction of propagation of the wave $v_{\phi}=\omega_r/k$. The following table summarize these properties:
\begin{center}
   \begin{tabular}{ | c | c | c | }
     \hline
      & $v_{\phi} > 0$ & $v_{\phi}< 0$  \\ \hline
     Positive helicity & Right polarization & Left polarization \\ \hline
     Negative helicity & Left polarization & Right polarization \\
     \hline
   \end{tabular}
\end{center}

\begin{center}
\section{Kinetic theory for a demagnetized plasma}
\label{sec:appendix_linear}
\end{center}
The dispersion relation for parallel propagating electromagnetic waves in plasma with Maxwellian populations can be written as:
\begin{align}
-k^2c^2-\frac{1}{\sqrt{2}}\sum_\alpha\left[\frac{\omega^2_{p\alpha}}{v_{T\alpha}}\left(u_{\parallel\alpha}-\frac{\omega}{k}\right)Z(\zeta^\pm_\alpha)\right]=0
\label{eq:appendix_dispersion}
\end{align}
where the thermal velocity is given by $v_{T\alpha}=(k_BT_{\alpha}/m_\alpha)^{1/2}$, $k_B$ is the Boltzmann constant, $\omega_{p\alpha}=(n_\alpha q_\alpha^2/\varepsilon_0 m_\alpha)^{1/2}$ is the plasma angular frequency, $\Omega_{\alpha}=q_\alpha B_0/m_\alpha$ is the initial cyclotron angular frequency, $\varepsilon_0$ is the permittivity of free space. The summation is performed over all populations $\alpha=e,m,cr$. We will restrict ourselves to low frequency waves, such that $\omega<\Omega_{\alpha}$. Under this assumption, $\zeta_\alpha^\pm$ can be rewritten as:
\begin{align}
\zeta_\alpha^\pm\approx\ \frac{1}{\sqrt{2}}\left(\frac{p^\pm}{kr_{L\alpha}}-\frac{u_{\parallel\alpha}}{v_{T\alpha}}\right)
\label{eq:zeta}
\end{align}
Eq. \ref{eq:zeta} highlights two different physical parameters. The first term is the ratio between the wavelength and the thermal Larmor radius, and characterizes the magnetization of the population. The second term is the ratio between the drift and thermal velocity of the population, which describes the velocity distribution. Populations with small thermal velocity and/or large drift velocity such as the cosmic rays will interact non-resonantly with the perturbations at a scale $k$, whereas sufficiently hot and slowly drifting population such as the main protons may interact resonantly or become demagnetized.

We will use the reference frame of the main protons. By making the assumption of magnetized electrons, demagnetized main protons, and of cosmic rays with a large drift over thermal velocity ratio, the arguments of the Fried and Conte functions in Eq. \ref{eq:appendix_dispersion} follow the limits: $|\zeta^\pm_{cr}|\gg 1$, $|\zeta^\pm_e|\gg 1$ and $|\zeta^\pm_m|< 1$. Using the appropriate asymptotic expansions, the Fried and Conte functions can then be rewritten as:
\begin{align}
Z(\zeta_{cr}^\pm)=&\ -\sqrt{2}\left(\frac{p^\pm}{kr_{Lcr}}-\frac{u_{\parallel cr}}{v_{Tcr}}\right)^{-1} + O(\zeta_{cr}^\pm)^3 \\
Z(\zeta_e^\pm)=&\ -\sqrt{2}\left(\frac{p^\pm}{kr_{Le}}-\frac{u_{\parallel e}}{v_{Te}}\right)^{-1} + O(\zeta_e^\pm)^3 \\
Z(\zeta_m^\pm)=&\ -\sqrt{2}\left(\frac{p^\pm}{kr_{Lm}}\right)+i\pi^{1/2}\ \ + O(\zeta_m^\pm)^3 
\end{align}
We have simplified the exponential terms and neglected the contributions of order $O(\zeta_\alpha^\pm)^3$. In the following, we will write $R_\alpha$ as the real part of the expansions for each populations $\alpha$. Inserting the Fried and Conte expansions in the dispersion relation gives:
\begin{align}
\begin{split}
\sqrt{2}k^2\frac{v_{A0}^2}{\Omega^2_{0}}=\frac{1}{v_{Tm}}\left[\frac{\omega}{k}(R_m+i\pi^{1/2})\right]&+\frac{R_e}{v_{Te}}\frac{\omega_{pe}^2}{\omega_{pm}^2}\left[\frac{\omega}{k}-u_{\parallel e}\right]+\frac{R_{cr}}{v_{Tcr}}\frac{\omega_{pcr}^2}{\omega_{pm}^2}\left[\frac{\omega}{k}-u_{\parallel cr}\right]
\end{split}
\end{align}
Separating the real and imaginary parts of $\omega= \omega_r+i\gamma$, one obtains:
\begin{align}
\gamma(k) = -k\dfrac{\left[\sqrt{2}k^2\dfrac{v_{A0}^2}{\Omega_{0}^2}+\dfrac{u_{\parallel e}}{v_{Te}}\dfrac{\omega_{pe}^2}{\omega_{pm}^2}R_e+\dfrac{u_{\parallel cr}}{v_{Tcr}}\dfrac{\omega_{pcr}^2}{\omega_{pm}^2}R_{cr}\right]\dfrac{\pi^{1/2}}{v_{Tm}}}{\left[\dfrac{R_m}{v_{Tm}}+\dfrac{R_e}{v_{Te}}\dfrac{\omega_{pe}^2}{\omega_{pm}^2}+\dfrac{R_{cr}}{v_{Tcr}}\dfrac{\omega_{pcr}^2}{\omega_{pm}^2}\right]^2+\dfrac{\pi}{v_{Tm}^2}}
\end{align}
\begin{align}
\omega_r(k) = k\dfrac{\left[\sqrt{2}k^2\dfrac{v_{A0}^2}{\Omega_{0}^2}+\dfrac{u_{\parallel e}}{v_{Te}}\dfrac{\omega_{pe}^2}{\omega_{pm}^2}R_e+\dfrac{u_{\parallel cr}}{v_{Tcr}}\dfrac{\omega_{pcr}^2}{\omega_{pm}^2}R_{cr}\right]\left[\dfrac{R_m}{v_{Tm}}+\dfrac{R_e}{v_{Te}}\dfrac{\omega_{pe}^2}{\omega_{pm}^2}+\dfrac{R_{cr}}{v_{Tcr}}\dfrac{\omega_{pcr}^2}{\omega_{pm}^2}\right]}{\left[\dfrac{R_m}{v_{Tm}}+\dfrac{R_e}{v_{Te}}\dfrac{\omega_{pe}^2}{\omega_{pm}^2}+\dfrac{R_{cr}}{v_{Tcr}}\dfrac{\omega_{pcr}^2}{\omega_{pm}^2}\right]^2+\dfrac{\pi}{v_{Tm}^2}}
\end{align}
We consider protons populations with a small density ratio $n_{cr}/n_m$, neglect electron inertia (which is equivalent to the low frequency assumption) and use the current condition (Eq \ref{eq:u_e}). After some algebra, one obtains the growth rate and real angular frequency in the hot, demagnetized regime of interaction:
\begin{align}
\begin{split}
\gamma_{\ms{hot}}(k) = &\dfrac{(2\pi)^{1/2}}{r_{Lm}\xi}
\dfrac{\dfrac{k}{\Omega_0}\biggl(v_{A0}^2-\dfrac{n_{cr}}{n_m}u_{\parallel cr}^2\biggr)-p^\pm\biggl(\dfrac{k^2}{\Omega_0^2}v_{A0}^2+\dfrac{n_{cr}^2}{n_m^2}\biggr)u_{\parallel cr}}{\dfrac{\pi}{k^2r_{Lm}^2}+2\biggl(\dfrac{1}{k^2r_{Lm}^2}-\dfrac{n_{cr}}{n_m}\dfrac{1}{\xi}-1\biggr)^2}
\end{split}
\end{align}
\begin{align}
\omega_{\ms{r,hot}}(k) = \dfrac{k^3r_{Lm}^2\left(k^2r_{Lm}^2-1\right)\left(\dfrac{n_{cr}}{n_m}u_{\parallel cr}+p^\pm \dfrac{k}{\Omega_0}v_{A0}^2\right)}{k^4r_{Lm}^4+k^2r_{Lm}^2\left(\dfrac{\pi}{2}-2\right)+1}
\end{align}
where we have defined the parameter $\xi=p^\pm ku_{\parallel cr}/\Omega_0-1$. We consider $kr_{Lm}\gg 1$ and $ku_{\parallel cr}/\Omega_0\gg 1$, which corresponds to the hypothesis of demagnetized main protons, and to the instability requirement $k>k_{min}$ discussed in Sec. \ref{sec:mechanism}. One finds:
\begin{align}
\gamma_{\ms{hot}}(k) = \left(\frac{\pi}{2}\right)^{1/2}\frac{1}{r_{Lm}u_{\parallel cr}}\left[p^\pm\left(v_{A0}^2-\frac{n_{cr}}{n_m}u_{\parallel cr}^2\right)-\left(k\frac{v_{A0}^2}{\Omega_0^2}+\frac{1}{k}\frac{n_{cr}^2}{n_m^2}\right)u_{\parallel cr}\Omega_0\right]
\end{align}
\begin{align}
\omega_{\ms{r,hot}}(k) = k\left(\frac{n_{cr}}{n_m}u_{\parallel cr}+p^\pm\frac{k}{\Omega_0}v_{A0}^2\right)
\end{align}
Calculating the growth rate derivative over $k$ and searching for an extremum yields:
\begin{align}
k_{\ms hot}=\frac{n_{cr}}{n_m}\frac{\Omega_0}{v_{A0}}
\end{align}
Inserting in the expressions of $\gamma_{\ms{hot}}(k)$ and $\omega_{\ms{r,hot}}(k)$, we obtain the growth rate, real angular frequency and phase velocity $v_{\phi,hot}=\omega_{\ms{r,hot}}/k_{\ms{hot}}$ for the fastest growing unstable mode:
\begin{align}
\gamma_{\ms{hot}} &= \left(\frac{\pi}{2}\right)^{1/2}\frac{n_{cr}}{n_m}\frac{u_{\parallel cr}}{v_{Tm}}\Omega_0 \\
\omega_{\ms{r,hot}} &= \frac{n_{cr}^2}{n_m^2}\frac{u_{\parallel cr}}{v_{A0}}\Omega_0  \\
v_{\phi,\ms hot}&=-\frac{n_{cr}}{n_m}u_{\parallel cr}
\end{align}

\bsp	
\label{lastpage}
\end{document}